\documentclass[12pt]{article}
\usepackage{jheppub}

\pdfoutput=1

\usepackage{amsmath,bbm,array,amsfonts,graphicx,wrapfig,lscape,float,mathtools,multirow,longtable}
\usepackage[dvipsnames]{xcolor}

\newcommand{\be}{\begin{equation}}
\newcommand{\ee}{\end{equation}}
\newcommand{\beq}{\begin{equation}}
\newcommand{\beql}[1]{\begin{equation}\label{#1}}
\newcommand{\eeq}{\end{equation}}
\newcommand{\ba}{\begin{array}}
\newcommand{\ea}{\end{array}}
\newcommand{\bea}{\begin{eqnarray}}
\newcommand{\beal}[1]{\begin{eqnarray}\label{#1}}
\newcommand{\eea}{\end{eqnarray}}
\newcommand{\ben}{\begin{enumerate}}
\newcommand{\een}{\end{enumerate}}
\newcommand{\bean}{\begin{eqnarray*}}
\newcommand{\eean}{\end{eqnarray*}}
\newcommand{\eref}[1]{(\ref{#1})}
\newcommand{\sref}[1]{\S\ref{#1}}
\newcommand{\tref}[1]{Table~\ref{#1}}

\newcommand{\fref}[1]{Figure \ref{#1}}
\newcommand{\btab}[1]{\begin{tabular}{#1}}
\newcommand{\etab}{\end{tabular}}

\def \Black {\pdfsetcolor{0 0 0 1}}

\newcommand{\comment}[1]{}

\newcommand{\qed}{\nobreak \ifvmode \relax \else
      \ifdim\lastskip<1.5em \hskip-\lastskip
      \hskip1.5em plus0em minus0.5em \fi \nobreak
      \vrule height0.75em width0.5em depth0.25em\fi}

\definecolor{darkspringgreen}{rgb}{0.09, 0.45, 0.27}
\definecolor{forestgreen}{rgb}{0.13, 0.55, 0.13}

\title{Orbifold Reduction and 2d (0,2) Gauge Theories}

\author[a,b]{Sebasti\'an Franco,} 
\author[c,d,e]{Sangmin Lee,}
\author[f]{Rak-Kyeong Seong}

\affiliation[a]{
Physics Department, The City College of the CUNY \\
160 Convent Avenue, New York, NY 10031, USA}

\affiliation[b]{The Graduate School and University Center, The City University of New York  \\
365 Fifth Avenue, New York NY 10016, USA }

\affiliation[c]{
Center for Theoretical Physics, Seoul National University, Seoul 08826, Korea
}

\affiliation[d]{
Department of Physics and Astronomy, Seoul National University, Seoul 08826, Korea
}

\affiliation[e]{
College of Liberal Studies, Seoul National University, Seoul 08826, Korea
}

\affiliation[f]{
School of Physics, Korea Institute for Advanced Study, Seoul 02455, Korea
}

\emailAdd{sfranco@ccny.cuny.edu}
\emailAdd{sangmin@snu.ac.kr}
\emailAdd{rakkyeongseong@gmail.com}

\preprint{
\begin{flushright}
CCNY-HEP-16-09 \\
SNUTP-16-005 \\
KIAS-P16071
\end{flushright}
}

\abstract{We introduce Orbifold Reduction, a new method for generating $2d$ $(0,2)$ gauge theories associated to D1-branes probing singular toric Calabi-Yau 4-folds starting from $4d$ $\mathcal{N}=1$ gauge theories on D3-branes probing toric Calabi-Yau 3-folds. The new procedure generalizes dimensional reduction and orbifolding. In terms of T-dual configurations, it generates brane brick models starting from brane tilings. Orbifold reduction provides an agile approach for generating $2d$ $(0,2)$ theories with a brane realization. We present three practical applications of the new algorithm: the connection between $4d$ Seiberg duality and $2d$ triality, a combinatorial method for generating theories related by triality and a $2d$ $(0,2)$ generalization of the Klebanov-Witten mass deformation.
}

\begin{document}

\maketitle

\section{Introduction}

D-branes probing singular Calabi-Yau (CY) manifolds provide a fruitful framework for engineering quantum field theories in various dimensions. These setups often give rise to new perspectives and powerful tools for understanding the dynamics of the corresponding gauge theories. 

A general program for studying the $2d$ $(0,2)$ gauge theories that live on the worldvolume of D1-branes probing singular toric CY 4-folds was recently initiated.\footnote{For earlier attempts at this question see \cite{Mohri:1997ef,GarciaCompean:1998kh}. Alternative approaches for realizing $2d$ $(0,2)$ theories in terms of branes can be found in \cite{Tatar:2015sga,Benini:2015bwz, Schafer-Nameki:2016cfr, Apruzzi:2016iac}.} In \cite{Franco:2015tna}, a systematic procedure for obtaining the $2d$ gauge theories on D1-branes probing  generic toric CY 4-folds was developed and the general properties of the theories arising from these setups were established. A new class of Type IIA brane configurations, denoted {\it brane brick models}, was introduced in \cite{Franco:2015tya}. Two of their most remarkable features are that they fully encode the gauge theories on the D1-branes and streamline the connection to the probed CY 4-folds. Brane brick models are related to D1-branes at singularities by T-duality. In \cite{Gadde:2013lxa}, a new order-3 IR equivalence among $2d$ $(0,2)$ theories, called {\it triality}, was discovered. The brane brick model realization of triality was investigated in \cite{Franco:2016nwv}. In \cite{Franco:2016qxh}, following \cite{Feng:2005gw,Futaki:2014mpa}, mirror symmetry was used to refine our understanding of the correspondence between D1-brane at singularities, brane brick models and $2d$ gauge theories. That work also explained how triality is realized in terms of geometric transitions in the mirror geometry.

In this paper we will introduce {\it orbifold reduction}, a new method for generating $2d$ $(0,2)$ gauge theories associated to D1-branes probing singular toric Calabi-Yau 4-folds starting from $4d$ $\mathcal{N}=1$ gauge theories on D3-branes probing toric Calabi-Yau 3-folds. This procedure generalizes dimensional reduction and orbifolding. Orbifold reduction allows us to generate the gauge theories for D1-branes probing complicated CY 4-folds with little effort. This feature makes it a powerful new addition to the toolkit for studying $2d$ $(0,2)$ theories in terms of D-branes and we consequently expect it will find several interesting applications.

This paper is organized as follows. In section \sref{section_2d_CY4_BBMs}, we review $2d$ $(0,2)$ theories, D1-branes over toric CY 4-folds and brane brick models. In section \sref{section_orbifold_reduction_general}, we introduce orbifold reduction. Section \sref{section_examples} contains explicit examples illustrating the construction. In order to demonstrate the usefulness of orbifold reduction we then present three possible applications. In \sref{section_triality_from_Seiberg}, we show how it generates $2d$ triality duals starting from $4d$ Seiberg dual theories. A distinctive feature of this approach is that both the $2d$ theories and their $4d$ parents are realized in terms of D-branes at singularities. In \sref{section_triality_combinatorics}, we show how the combinatorics of orbifold reduction leads to non-trivial triality duals. Finally, in section \sref{section_KW}, we use orbifold reduction to construct an explicit example of a theory admitting a $2d$ $(0, 2)$ generalization of the Klebanov-Witten mass deformation. We conclude in section \sref{section_conclusions}. Additional examples are presented in appendix \sref{section_appendix_examples}.

\section{2d (0,2) Theories, Toric CY$_4$'s and Brane Brick Models}

\label{section_2d_CY4_BBMs}

The study of the $2d$ $(0,2)$ gauge theories that arise on the worldvolume of D1-branes probing toric CY 4-folds was developed in \cite{Franco:2015tna,Franco:2015tya,Franco:2016nwv,Franco:2016qxh}, to which we refer the reader for details. The probed CY 4-fold arises as the classical mesonic moduli space of the gauge theory on the D1-branes.

{\it Brane brick models} are Type IIA brane configurations that are related to D1-branes at toric singularities by T-duality. They were introduced in \cite{Franco:2015tna,Franco:2015tya} and they considerably simplify the connection between gauge theory and the probed CY 4-fold. 

A brane brick model consists of D4-branes suspended from an NS5-brane as summarized in \tref{tbconfig}. The $(01)$ directions, which are common to all the branes, support the $2d$ $(0,2)$ gauge theory. The NS5-brane also wraps a holomorphic surface $\Sigma$ embedded in $(234567)$. The coordinates $(23)$, $(45)$ and $(67)$ form three complex variables $x$, $y$ and $z$. The arguments of these variables are identified with $(246)$, which hence form a $T^3$. The surface $\Sigma$ is the zero locus of the Newton polynomial associated to the toric diagram of the CY$_4$, $P(x,y,z)=0$. Stacks of D4-branes extend along $(246)$ and are suspended from the NS5-brane. 

\begin{table}[ht!!]
\centering
\begin{tabular}{c|cccccccccc}
\; & 0 & 1 & 2 & 3 & 4 & 5 & 6 & 7 & 8 & 9\\
\hline
\text{D4} & $\times$ & $\times$ & $\times$ & $\cdot$ & $\times$ & $\cdot$ & $\times$ & $\cdot$ & $\cdot$ & $\cdot$ \\
\text{NS5} & $\times$ & $\times$ & \multicolumn{6}{c}{----------- \ $\Sigma$ \ ------------} & $\cdot$ & $\cdot$\\
\end{tabular}
\caption{Brane brick models are Type IIA configurations with D4-branes suspended from an NS5-brane that wraps a holomorphic surface $\Sigma$.}
\label{tbconfig}
\end{table}

It is convenient to represent a brane brick model by its ``skeleton'' on $T^3$. A brane brick model fully encodes a $2d$ $(0,2)$ gauge theory following the dictionary in \tref{tbrick}. Bricks correspond to $U(N)$ gauge groups.\footnote{It is possible to have different ranks by introducing fractional D1-branes in the T-dual configuration of branes at a CY$_4$ singularity.} There are two types of faces: oriented and unoriented faces correspond to chiral and Fermi fields, respectively.\footnote{We refer the reader to \cite{Franco:2015tya,Franco:2016qxh} for discussions on how to systematically orient faces.} We will identify chiral and Fermi faces by coloring them grey and red, respectively. Every edge in the brane brick model is attached to a single face corresponding to a Fermi field and a collection of faces corresponding to chiral fields. The chiral faces attached to an edge form a holomorphic monomial product that corresponds to either a $J$- or $E$-term that is associated to the Fermi field attached to the same edge.\footnote{It is possible to have non-generic gauge theory phases that correspond to brane brick models in which two Fermi faces share a common edge. In such cases, the $J$- and $E$-terms can be determined using alternative methods, such as partial resolution or triality \cite{Franco:2015tna,Franco:2015tya,Franco:2016nwv}.} Fermi faces are always 4-sided. This follows from the special structure of $J$- and $E$-terms in toric theories, which are always of the form 
\beal{esb1a1}
J_{ji} = J_{ji}^{+} - J_{ji}^{-} ~,~ E_{ij} = E_{ij}^{+} - E_{ij}^{-} ~,~
\eea
with $J_{ji}^{\pm}$ and $E_{ij}^{\pm}$ holomorphic monomials in chiral fields \cite{Franco:2015tna}.

\begin{table}[h]
\centering
\resizebox{0.9\hsize}{!}{
\begin{tabular}{|l|l|}
\hline
\ \ \ \ \ \ {\bf Brane Brick Model} \ \ \ \ \ & \ \ \ \ \ \ \ \ \ \ \ \ \ \ \ \ \ \ \ \ {\bf Gauge Theory} \ \ \ \ \ \ \ \ \ \ \ \ 
\\
\hline\hline
Brick  & Gauge group \\
\hline
Oriented face between bricks & Chiral field in the bifundamental representation \\
$i$ and $j$ & of nodes $i$ and $j$ (adjoint for $i=j$) \\
\hline
Unoriented square face between & Fermi field in the bifundamental representation \\
bricks $i$ and $j$ & of nodes $i$ and $j$ (adjoint for $i=j$) \\
\hline
Edge  & Plaquette encoding a monomial in a \\ 
& $J$- or $E$-term \\
\hline
\end{tabular}
}
\caption{
Dictionary between brane brick models and $2d$ $(0,2)$ gauge theories.
\label{tbrick}
}
\end{table}

\begin{figure}[ht]
\begin{center}
\resizebox{0.7\hsize}{!}{
\includegraphics[width=8cm]{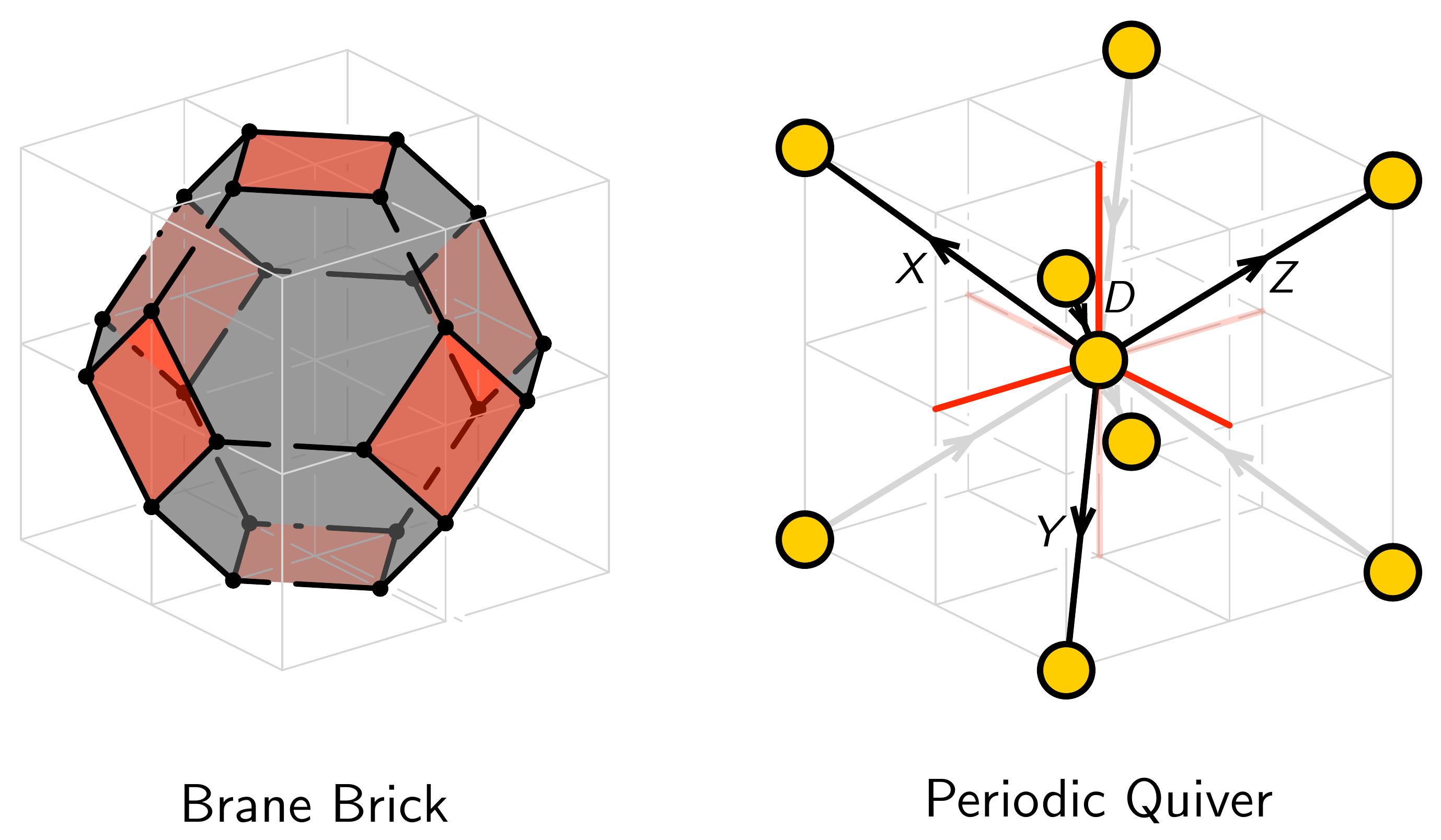}
}
\caption{A truncated octahedron as the brane brick for the brane brick model corresponding to $\mathbb{C}^4$ and the corresponding periodic quiver on $T^3$. In brane brick models, chiral and Fermi fields are represented by grey oriented faces and red unoriented square faces, respectively. In periodic quivers, chiral and Fermi fields correspond to arrows and unoriented edges, respectively.
\label{BBM_and_periodic_quiver}}
 \end{center}
 \end{figure}

Brane brick models are in one-to-one correspondence with {\it periodic quivers} on $T^3$ \cite{GarciaCompean:1998kh,Franco:2015tna,Franco:2015tya}. The two types of objects are related by graph dualization as shown in \fref{BBM_and_periodic_quiver}. Then, periodic quivers also uniquely define a $2d$ $(0,2)$ gauge theory. In particular, $J$- and $E$-terms correspond to minimal plaquettes. A plaquette is a gauge invariant closed loop in the quiver consisting of an oriented path of chiral fields and a single Fermi field. Most of our discussion in this paper will be phrased in terms of periodic quivers. This choice is motivated by the relative simplicity of the resulting figures. Constructing the corresponding brane brick models is straightforward.

\section{Orbifold Reduction}

\label{section_orbifold_reduction_general}

In this section we introduce orbifold reduction. It is a natural generalization of dimensional reduction and orbifolding, so we review them first.

\subsection{Dimensional Reduction}

Let us start with the dimensional reduction of general $4d$ $\mathcal{N}=1$ theories down to $2d$ $(2,2)$ theories. Under dimensional reduction, the $4d$ vector $\mathcal{V}_i$ and chiral $\mathcal{X}_{ij}$ multiplets become $2d$ $(2,2)$ vector and chiral multiplets, respectively. In terms of $2d$ $(0,2)$ multiplets, we have:

\begin{itemize}
\item \underline{$4d$ $\mathcal{N}=1$ vector $\mathcal{V}_i$} $\rightarrow$ $2d$ $(0,2)$ vector $V_i$ + $2d$ $(0,2)$ adjoint chiral $\Phi_{ii}$
\item \underline{$4d$ $\mathcal{N}=1$ chiral $\mathcal{X}_{ij}$} $\rightarrow$ $2d$ $(0,2)$ chiral $X_{ij}$ + $2d$ $(0,2)$ Fermi $\Lambda_{ij}$
\end{itemize}  

\noindent The $J$- and $E$-terms of the $2d$ theory are
\beal{es100}
J_{ji} = \frac{\partial W}{\partial X_{ij}} \, , \qquad
E_{ij} = \Phi_{ii} X_{ij} - X_{ij} \Phi_{jj}\, ,
\eea
where $W$ is the superpotential of the $4d$ theory.

\subsection*{Dimensional Reduction of Toric Theories}

Let us now focus on the class of $4d$ $\mathcal{N}=1$ theories that arise on D3-branes probing toric CY$_3$ singularities. Such theories are fully encoded by brane tilings, which are bipartite graphs on $T^2$ (see \cite{Hanany:2005ve,Franco:2005rj} for details). GLSM fields in the toric description of the CY$_3$, namely points in its toric diagram, admit a combinatorial implementation as perfect matchings of the brane tiling. A perfect matching $p_\alpha$ is a collection of edges in the tiling such that every node is the endpoint of exactly one edge in $p_\alpha$. Given the map between brane tilings and $4d$ gauge theories, we can regard perfect matchings as collections of chiral fields. In general, more than one perfect matching is associated to a given point in the toric diagram. This will become important later when we discuss orbifold reduction. For details on how to determine perfect matchings and connect them to toric diagrams we refer the reader to \cite{Hanany:2005ve} (see also \cite{Franco:2012mm} for a more modern perspective).

Brane tilings are in one-to-one correspondence with periodic quivers on $T^2$ via graph dualization. These periodic quivers are equivalent to brane tilings, so they fully specify the corresponding gauge theories. In particular, every plaquette in the periodic quiver corresponds to a term in the superpotential. 

\begin{figure}[ht]
\begin{center}
\resizebox{0.8\hsize}{!}{
\includegraphics[width=8cm]{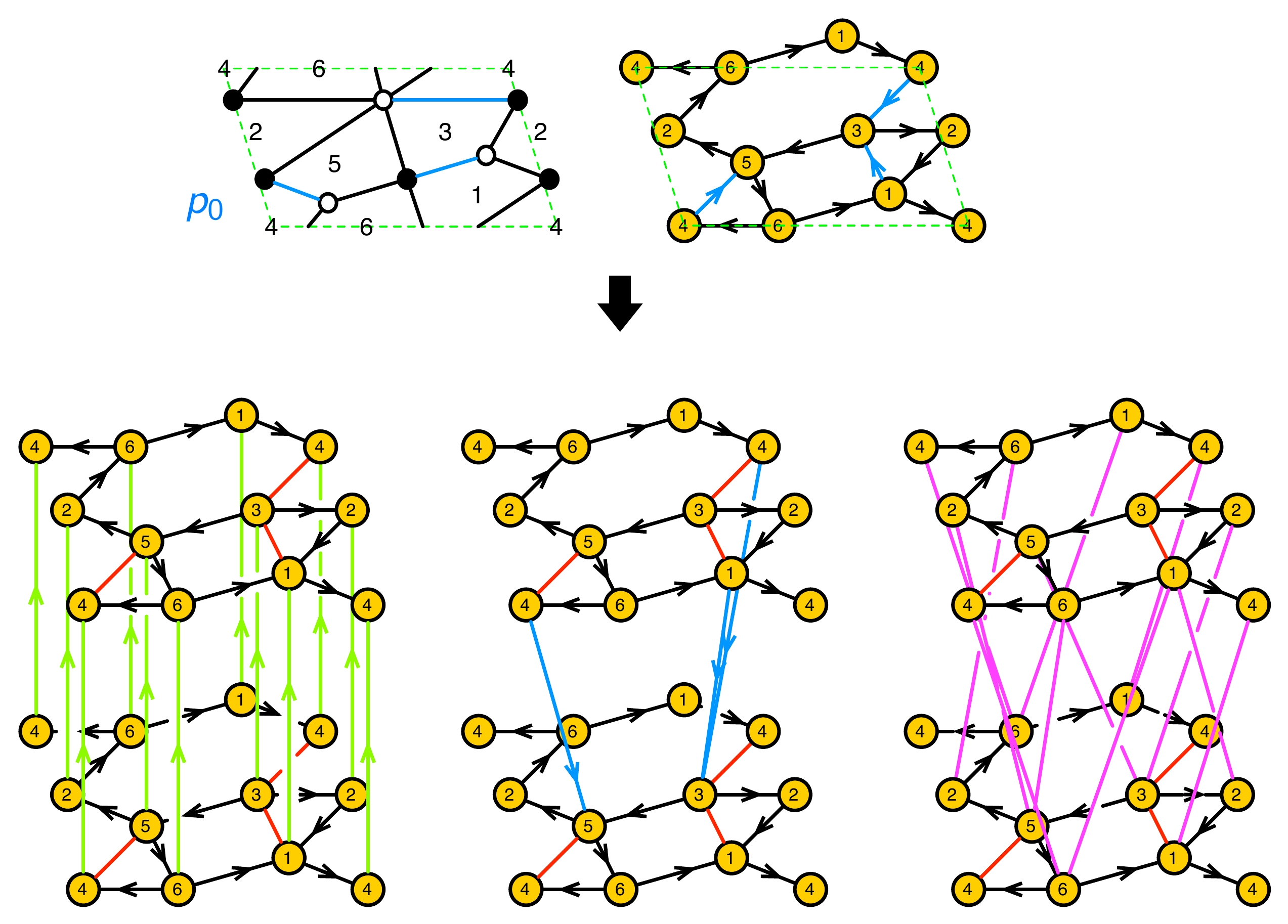}
}
\caption{Dimensional reduction of phase 1 of $dP_3$ using the perfect matching $p_0 = \{X_{45},X_{13},X_{43}\}$.
\label{dimred_dp3}}
 \end{center}
 \end{figure}

The dimensional reduction of a $4d$ $\mathcal{N}=1$ theory associated to a toric CY$_3$ gives rise to a $2d$ $(2,2)$ theory corresponding to the toric $\mathrm{CY}_4 = \mathrm{CY}_3 \times \mathbb{C}$. A systematic {\it lifting algorithm} for constructing the periodic quiver on $T^3$ associated to $\mathrm{CY}_3 \times \mathbb{C}$
starting from the periodic quiver on $T^2$ for $\mathrm{CY}_3$ was introduced in \cite{Franco:2015tna}. Let us refer to the new periodic direction of $T^3$ as the {\it vertical} direction. The lift is achieved by providing chiral and Fermi fields with vertical shifts. The vertical shifts for chiral fields are measured between the tail and the head of arrows. For Fermi fields, we use the same prescription, with the orientation dictated by the corresponding $4d$ chiral field. The procedure has a beautiful combinatorial implementation in terms of perfect matchings of the original brane tiling. For any perfect matching $p_0$, the periodic quiver for the dimensionally reduced theory is achieved by introducing the following vertical shifts for the different types of matter fields:
\beq
\Phi_{ii} \to 1 \qquad, \qquad X_{ij}  \to  \begin{array}{rl} -1 & \mbox{if } \chi_{ij} \in p_0 \\[.1cm]
 0 & \mbox{if } \chi_{ij} \notin p_0 
\end{array} \qquad , \qquad
\Lambda_{ij}  \to  \begin{array}{rl} 0 & \mbox{if } \chi_{ij} \in p_0 \\[.1cm]
1 & \mbox{if } \chi_{ij} \notin p_0 
\end{array} \, ,
\label{rule_vertical_shift_1}
\eeq
where 1 is the periodicity of the vertical direction. As it is standard in the study of brane tilings \cite{Franco:2005rj}, let us introduce $\langle \chi_{ij},p_0 \rangle$, which is defined to be equal to 1 if  $\chi_{ij} \in p_0$ and 0 if $\chi_{ij} \notin p_0$. The vertical shifts can then be compactly written as
\beq
\Phi_{ii} \to 1 \qquad , \qquad X_{ij}  \to  -\langle \chi_{ij},p_0 \rangle \qquad , \qquad \Lambda_{ij}  \to  \langle \chi_{ij},p_0 \rangle -1 \, .
\label{rule_vertical_shift_2}
\eeq
\fref{dimred_dp3} illustrates this procedure with the dimensional reduction of phase 1 of $dP_3$ \cite{Feng:2002zw,Franco:2005rj}. Let us explain how to interpret this kind of figure. In order to avoid clutter, throughout this paper we will represent periodic quivers in terms of three figures. Each of them contains different types of fields between layers: chirals coming from $4d$ vector multiplets in green, chirals coming from $4d$ chirals in blue and Fermis coming from $4d$ chirals in magenta. The combination of the three figures should be regarded as a single periodic quiver. In addition, for clarity, we will frequently present a region of the quiver that is larger than a unit cell, which is easy to determine from the node labels.

\begin{figure}[ht]
\begin{center}
\resizebox{0.5\hsize}{!}{
\includegraphics[width=8cm]{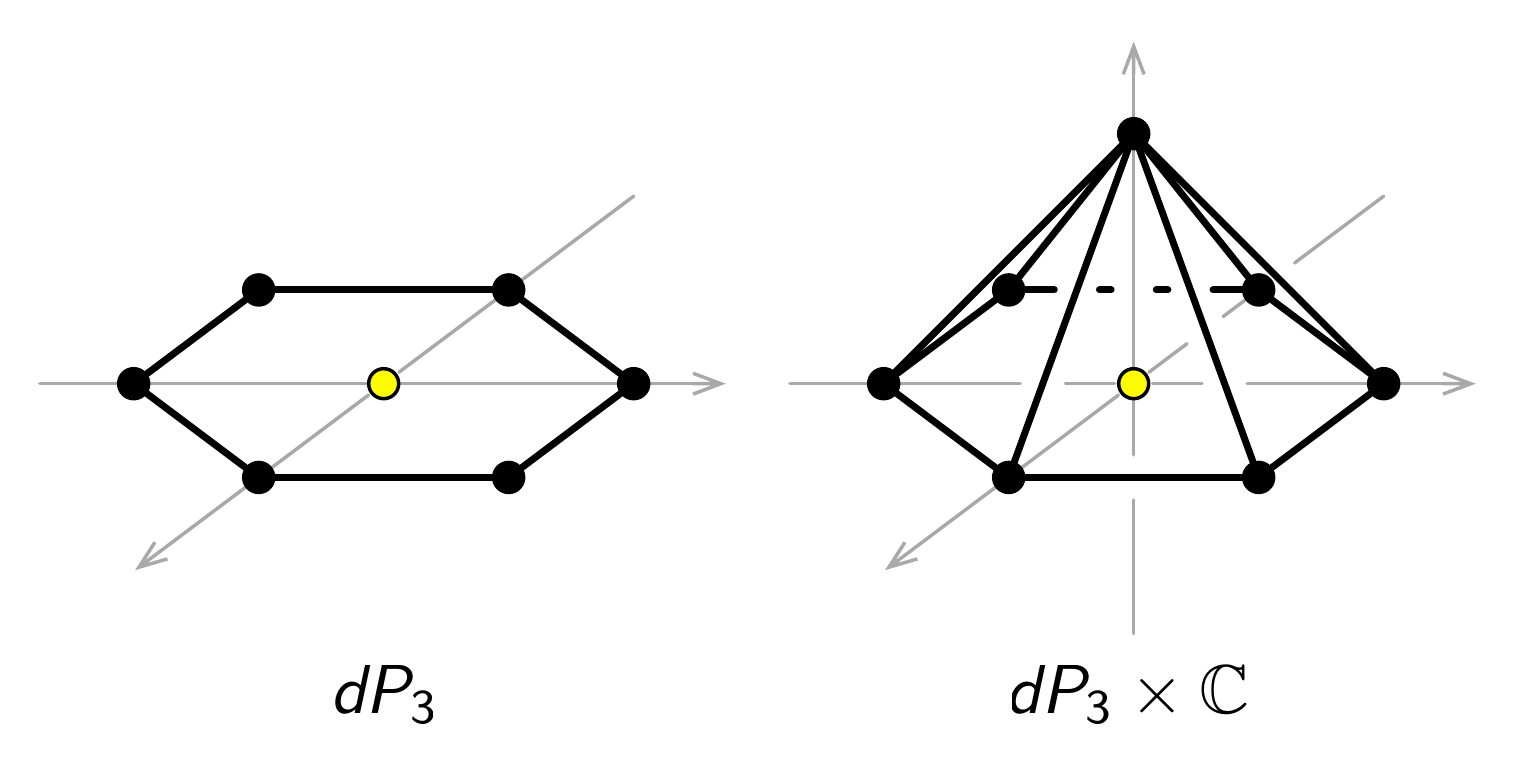}
}
\caption{The toric diagrams for $dP_3$ and $dP_3 \times \mathbb{C}$.
\label{dp3anddp3c}}
 \end{center}
 \end{figure}

Any perfect matching can be used to perform the dimensional reduction and the final result is independent of this choice. The $2d$ toric diagram of the CY$_3$ becomes the $3d$ toric diagram of the $\mathrm{CY}_4=\mathrm{CY}_3 \times \mathbb{C}$ by adding an extra point that represents the $\mathbb{C}$ factor, as shown in \fref{dp3anddp3c}. This additional point can be regarded as arising from $p_0$. It can be moved to any position on a plane parallel to the $2d$ toric diagram by an $SL(3,\mathbb{Z)}$ transformation, which is another indication of the freedom in choosing $p_0$.

If we dimensionally reduce two $4d$ theories connected by Seiberg duality, we obtain a pair of $2d$ $(2,2)$ theories related by the duality of \cite{Benini:2014mia}.

While we are not going to use it in this paper, it is worth mentioning that dimensional reduction also has a beautiful implementation as a lift of brane tilings into brane brick models \cite{Franco:2015tya}.

\subsection{Orbifolding}

In terms of brane brick models and periodic quivers, orbifolds are constructed by enlarging the unit cell \cite{Franco:2015tna,Franco:2015tya}. The details of the geometric action of the orbifold group are encoded in the periodic identifications on $T^3$. 

The lifting algorithm of the previous section can be extended to generate the gauge theories for certain $(\mathrm{CY}_3 \times \mathbb{C})/\mathbb{Z}_k$ orbifolds.\footnote{Not all possible orbifolds of $\mathrm{CY}_3 \times \mathbb{C}$ can be generated in this way. This is rather clear, since we are not considering all possible ways of enlarging the unit cell.} 
The process is very simple:

\begin{enumerate}
\item Stack $k$ copies of the gauge nodes in the $T^2$ periodic quiver along the vertical direction. We set the distance between consecutive layers equal to 1.
\item Choose a perfect matching $p_0$.
\item For each of the $k$ layers, introduce matter fields with vertical shifts given by \eref{rule_vertical_shift_2}.
\end{enumerate}

\begin{figure}[ht]
\begin{center}
\resizebox{0.6\hsize}{!}{
\includegraphics[width=8cm]{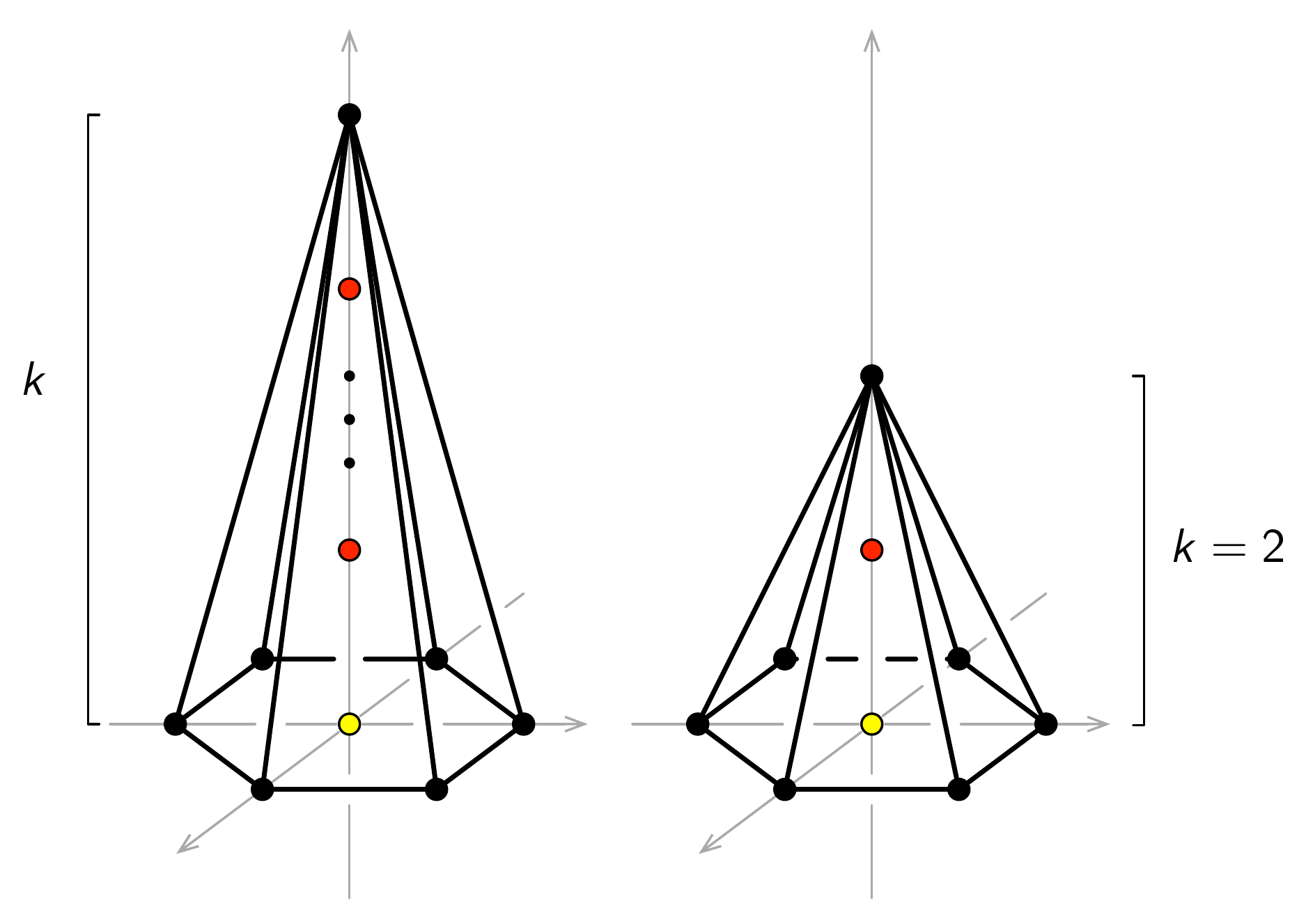}
}
\caption{The toric diagrams for certain orbifolds of the form $(dP_3 \times \mathbb{C})/\mathbb{Z}_k$, with $k$ controlling the height of the toric diagram.
\label{dp3czk}}
 \end{center}
 \end{figure}

The effect on the toric diagram is to expand the point associated to $p_0$ into a line of length $k$, as shown in \fref{dp3czk}. Once again, any perfect matching can be used in this construction. However, unlike in dimensional reduction, perfect matchings associated to different points in the toric diagram give rise to non-$SL(3,Z)$ equivalent geometries. In other words, they correspond to different actions of the orbifold group.

\subsection{Orbifold Reduction}

\label{section_orbifold_reduction}

Orbifold reduction is a simple generalization of the orbifolding procedure discussed in the previous section and can be summarized as follows:

\begin{enumerate}
\item Stack $k$ copies of the gauge nodes in the $T^2$ periodic quiver along the vertical direction. The distance between consecutive layers is set to 1.
\item Choose a perfect matching $p_0$ and a $k$-dimensional vector of signs $s=(s_1,\dots,s_k)$, with $s_i=\pm$.
\item Between layers $i$ and $i+1$, we introduce matter fields according to \eref{rule_vertical_shift_2} if $s_i=+$ or to its vertical reflection if $s_i=-$. Equivalently, if $s_i=-$, the vertical shifts of fields are given by minus \eref{rule_vertical_shift_2}, measured with respect to the $i+1$ layer.
\end{enumerate}

\begin{figure}[ht]
\begin{center}
\resizebox{0.35\hsize}{!}{
\includegraphics[width=8cm]{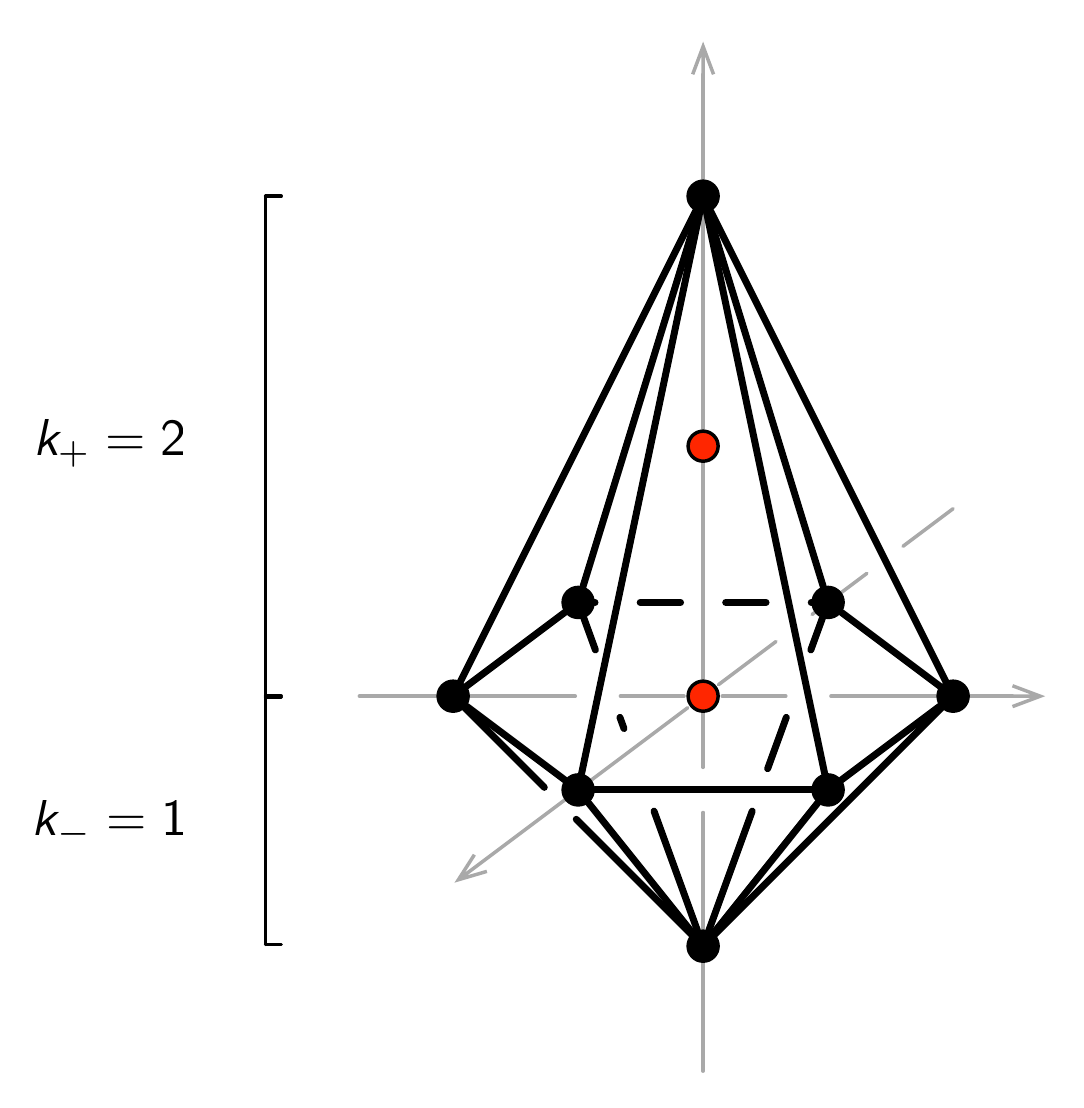}
}
\caption{Toric diagram for the orbifold reduction of $dP_3$ with $k_+=2$ and $k_-=1$.
\label{dp3czkex}}
 \end{center}
 \end{figure}

Let us denote $k_+$ and $k_-$ the number of plus and minus signs in $s$, respectively. Orbifold reduction generates a gauge theory that corresponds to a CY$_4$ whose toric diagram is obtained by expanding the point associated to $p_0$ into a line of length $k$, with $k_+$ points above the original $2d$ toric diagram and $k_-$ points below it. This generalizes the transformations of toric diagrams discussed in the two previous sections. \fref{dp3czkex} shows an example. 

Clearly $s$ and $-s$ give rise to the same theory, since their periodic quivers are simply related by a reflection along the vertical direction. Furthermore, $s=(+,\ldots,+)$ corresponds to a $(\mathrm{CY}_3 \times \mathbb{C})/\mathbb{Z}_k$ orbifold. In particular $s=(+)$ corresponds to dimensional reduction.

We previously saw that any perfect matching can be used as $p_0$ for both dimensional reduction and orbifolding. However, this is no longer the case for orbifold reduction. In particular, certain combinations of $s$ and $p_0$ can result in theories with non-vanishing non-abelian anomalies. Hence, it is always necessary to verify at an initial stage that we pick a $p_0$ that, in combination with $s$, does not give rise to anomalies. In order to determine whether a given $p_0$ is consistent with any $s$, it is sufficient to check that it does not lead to anomalies for a layer between $(+,-)$ signs (layers between equal signs are automatically free of anomalies, as in dimensional reduction and orbifolding). Our study suggests that it is always possible to pick a $p_0$ such that any point in a $2d$ toric diagram can be consistently lifted. It would be interesting to prove this fact.

Denoting $G^{(2d)}$, $N_\chi ^{(2d)}$ and $N_F^{(2d)}$ the numbers of gauge groups, chirals and Fermis in the orbifold reduced theory and $G^{(4d)}$ and $N_\chi ^{(4d)}$ the numbers of gauge groups and chirals in the $4d$ parent, we have
\beq
\begin{array}{ccl}
G^{(2d)} & = & k \,G^{(4d)} \, , \\[.15cm]
N_\chi ^{(2d)} & = & k \,N_\chi ^{(4d)} \, , \\[.15cm]
N_F^{(2d)} & = & k \,(G^{(4d)}+N_\chi ^{(4d)}) \, .
\end{array}
\label{2d_field_content_from_4d}
\eeq

A convenient notation for specifying a $2d$ theory $T_2$ obtained by orbifold reduction is
\beq
T_2=T_{4,s}(p_0) \, ,
\eeq
which emphasizes the necessary data for performing orbifold reduction: a toric $4d$ parent theory $T_4$, a sign vector $s$ and a perfect matching $p_0$. In this paper, we will not consider pairs of theories that differ only by the choice of $p_0$. For simplicity, we will thus omit $p_0$ from the label of $T_2$.

\newpage

\section{Examples}

\label{section_examples}

Below we illustrate orbifold reduction with two simple examples. They are the two inequivalent $k=2$ models obtained from the conifold: $\mathcal{C}_{(+,+)}$ and $\mathcal{C}_{(+,-)}$. Additional examples are presented in the appendix.

\subsection{$\mathcal{C}_{(+,+)}$}

\fref{conifoldquivertoric} shows the toric diagram and periodic quiver for the conifold $\mathcal{C}$. Each of the four points in the toric diagram corresponds to a single perfect matching. The four points are equivalent, so any choice of $p_0$ to be used in orbifold reduction generates the same theory. 

\begin{figure}[ht]
\begin{center}
\resizebox{0.6\hsize}{!}{
\includegraphics[width=8cm]{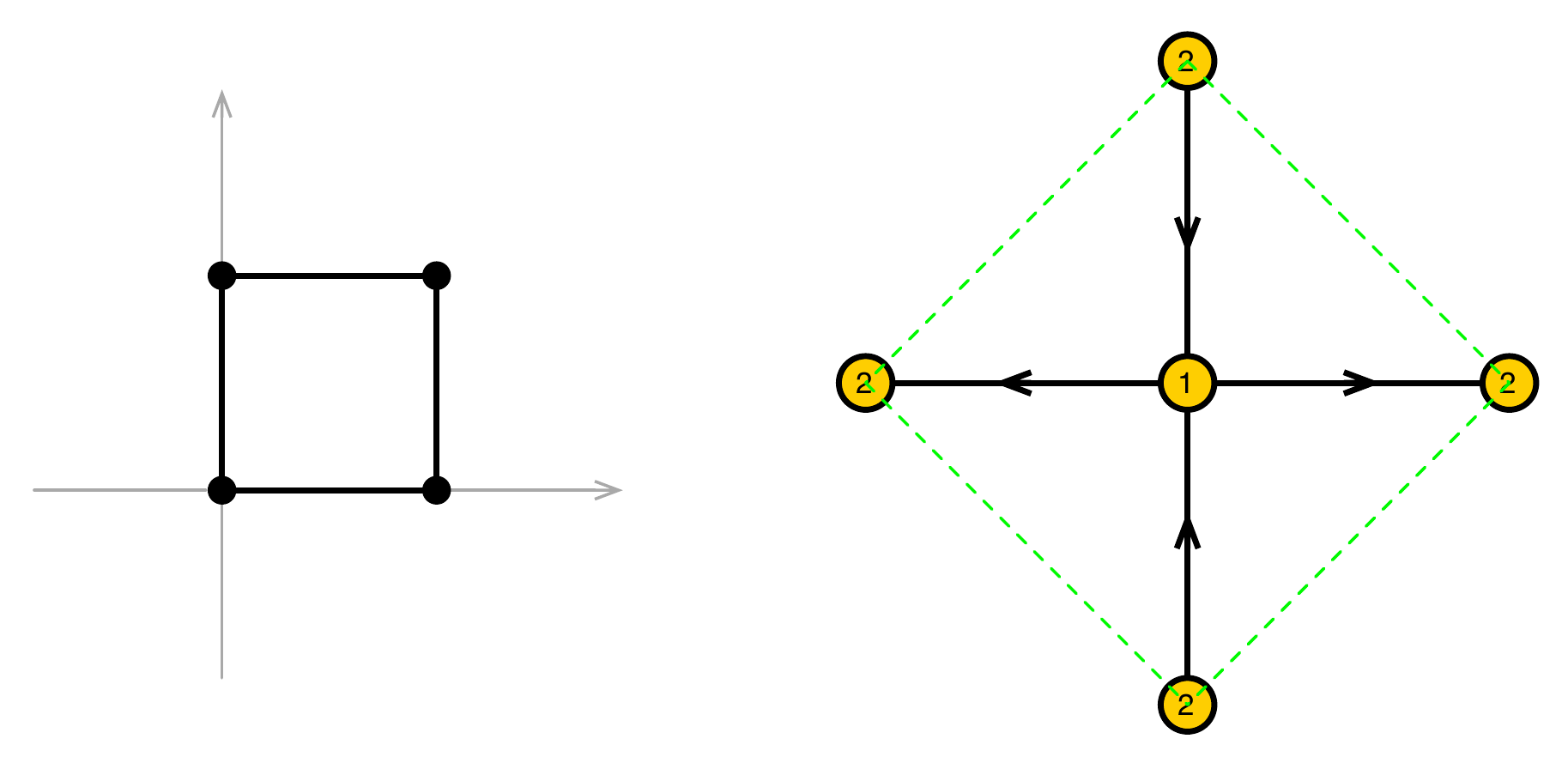}
}
\caption{Toric diagram and periodic quiver for the conifold $\mathcal{C}$.
\label{conifoldquivertoric}}
 \end{center}
 \end{figure} 

\begin{figure}[ht]
\begin{center}
\includegraphics[height=6.5cm]{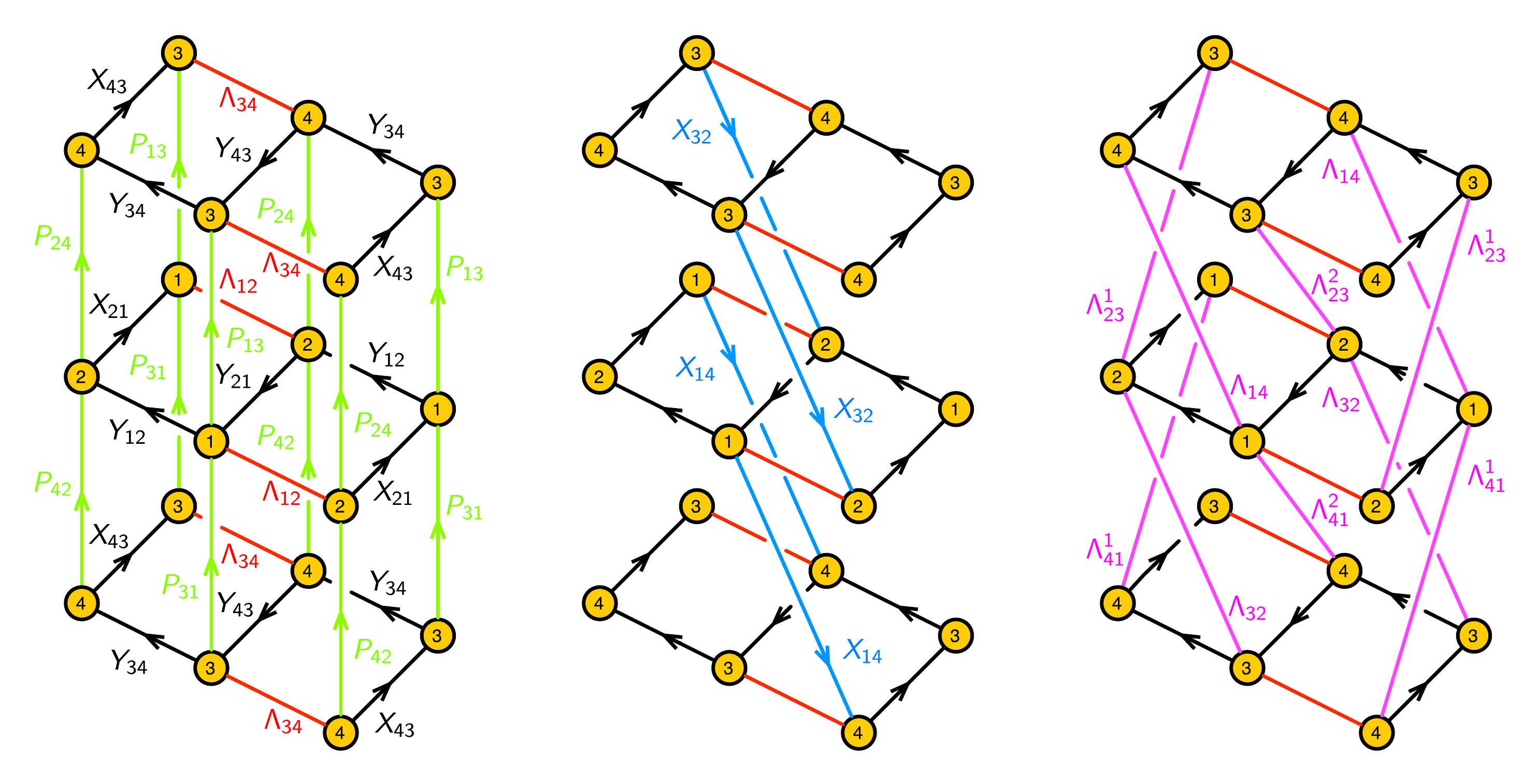}
\caption{Periodic quiver for $\mathcal{C}_{(+,+)}$.
\label{periodic_quiver_C++}}
 \end{center}
 \end{figure}

Performing orbifold reduction with $s=(+,+)$, we obtain a $2d$ theory that we call $\mathcal{C}_{(+,+)}$. Its periodic quiver is shown in \fref{periodic_quiver_C++}.
 
In all the examples we consider in this paper, it is straightforward to identify the $p_0$ that was used from the periodic quiver, as we now explain. The quiver on any of the layers is almost identical to the periodic quiver for the $4d$ parent. The only difference is given by the Fermi fields (red edges) which, according to \eref{rule_vertical_shift_2}, correspond precisely to the chiral fields contained in $p_0$. \fref{dimred_dp3} gives a good idea of how this works.

From the periodic quiver, we read the following $J$- and $E$- terms:
\beq
\begin{array}{rcrcccrcc}
& & J \ \ \ \ \ \ \ \ \ \ \ \ \ \ \ \ \ \ \ & & & & E \ \ \ \ \ \ \ \ \ \ \ \ & & \\
\Lambda_{12} : & \ \ \  & X_{21} \cdot Y_{12} \cdot Y_{21} - Y_{21} \cdot Y_{12} \cdot X_{21} & = & 0 & \ \ \ \  & P_{13} \cdot X_{32} - X_{14} \cdot P_{42} & = & 0 \\ 
\Lambda_{14} : & \ \ \  & Y_{43} \cdot X_{32} \cdot X_{21} - X_{43} \cdot X_{32} \cdot Y_{21} & = & 0 & \ \ \ \  & P_{13} \cdot Y_{34} - Y_{12} \cdot P_{24} & = & 0 \\
\Lambda_{32} : & \ \ \  & Y_{21} \cdot X_{14} \cdot X_{43} - X_{21} \cdot X_{14} \cdot Y_{43} & = & 0 & \ \ \ \  & P_{31} \cdot Y_{12} - Y_{34} \cdot P_{42} & = & 0 \\  
\Lambda_{34} : & \ \ \  & X_{43} \cdot Y_{34} \cdot Y_{43} - Y_{43} \cdot Y_{34} \cdot X_{43} & = & 0 & \ \ \ \  & P_{31} \cdot X_{14} - X_{32} \cdot P_{24} & = & 0 \\ 
\Lambda_{23}^{1} : & \ \ \  & Y_{34} \cdot Y_{43} \cdot X_{32} - X_{32} \cdot Y_{21} \cdot Y_{12} & = & 0 & \ \ \ \  & P_{24} \cdot X_{43} - X_{21} \cdot P_{13} & = & 0 \\ 
\Lambda_{23}^{2} : & \ \ \  & X_{32} \cdot X_{21} \cdot Y_{12} - Y_{34} \cdot X_{43} \cdot X_{32} & = & 0 & \ \ \ \  & P_{24} \cdot Y_{43} - Y_{21} \cdot P_{13} & = & 0 \\ 
 \end{array}
 \nonumber
\label{E_J_C_++}
 \eeq

\beq
\begin{array}{rcrcccrcc}
\Lambda_{41}^{1} : & \ \ \  & Y_{12} \cdot Y_{21} \cdot X_{14} - X_{14} \cdot Y_{43} \cdot Y_{34} & = & 0 & \ \ \ \  & P_{42} \cdot X_{21} - X_{43} \cdot P_{31} & = & 0 \\ 
\Lambda_{41}^{2} : & \ \ \  & X_{14} \cdot X_{43} \cdot Y_{34} - Y_{12} \cdot X_{21} \cdot X_{14} & = & 0 & \ \ \ \  & P_{42} \cdot Y_{21} - Y_{43} \cdot P_{31} & = & 0
 \end{array}
\label{E_J_C_++}
 \eeq
 
\begin{figure}[H]
\begin{center}
\includegraphics[height=6.5cm]{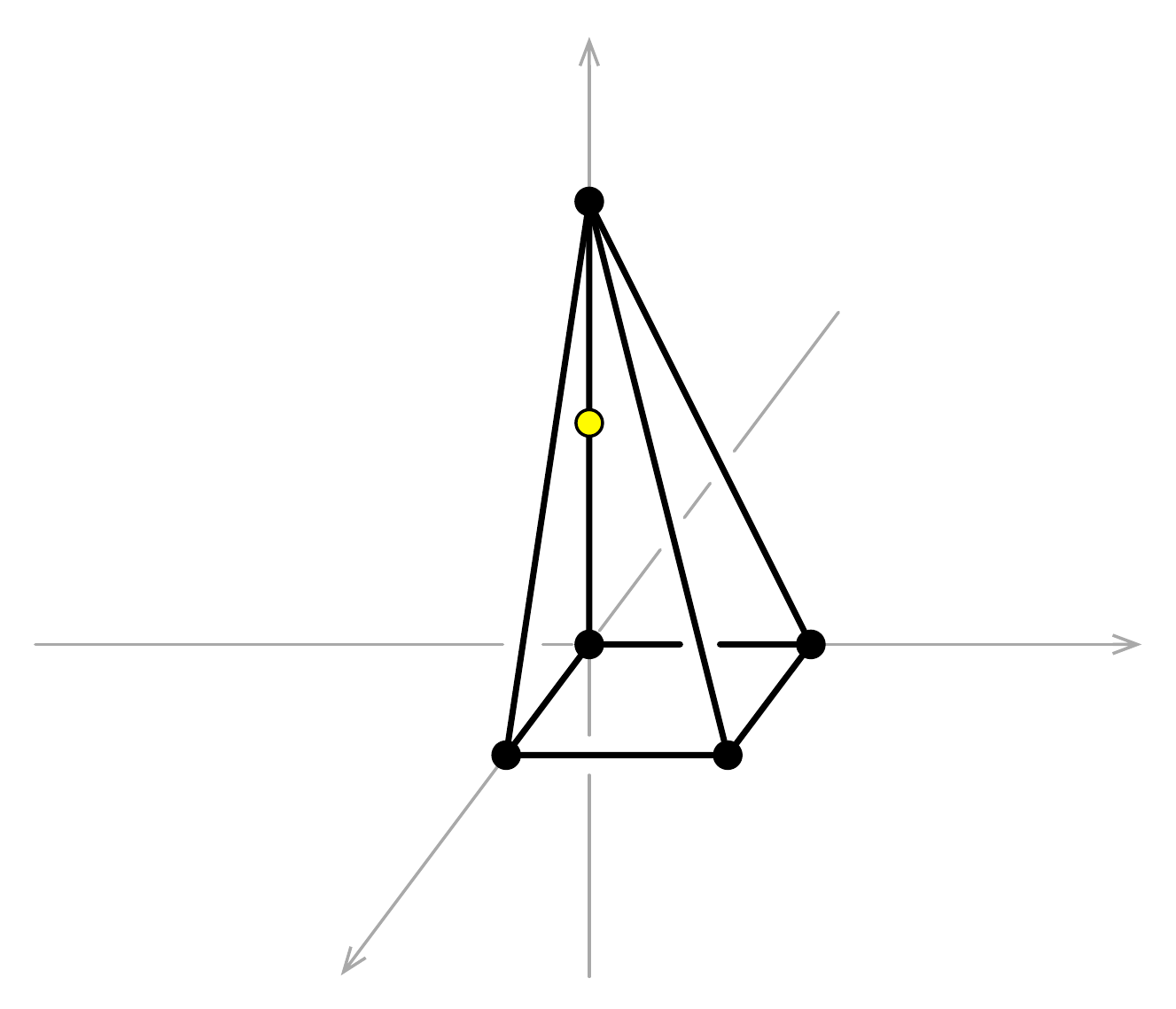}
\caption{Toric diagram for $\mathcal{C}_{(+,+)}$. 
}
\label{toric_C++}
 \end{center}
 \end{figure}

Computing the classical mesonic moduli space for this theory, we obtain the toric diagram that is expected from the general discussion in section \sref{section_orbifold_reduction}, which is shown in \fref{toric_C++}. 
 
\subsection{$\mathcal{C}_{(+,-)}$}

\begin{figure}[H]
\begin{center}
\includegraphics[height=6.5cm]{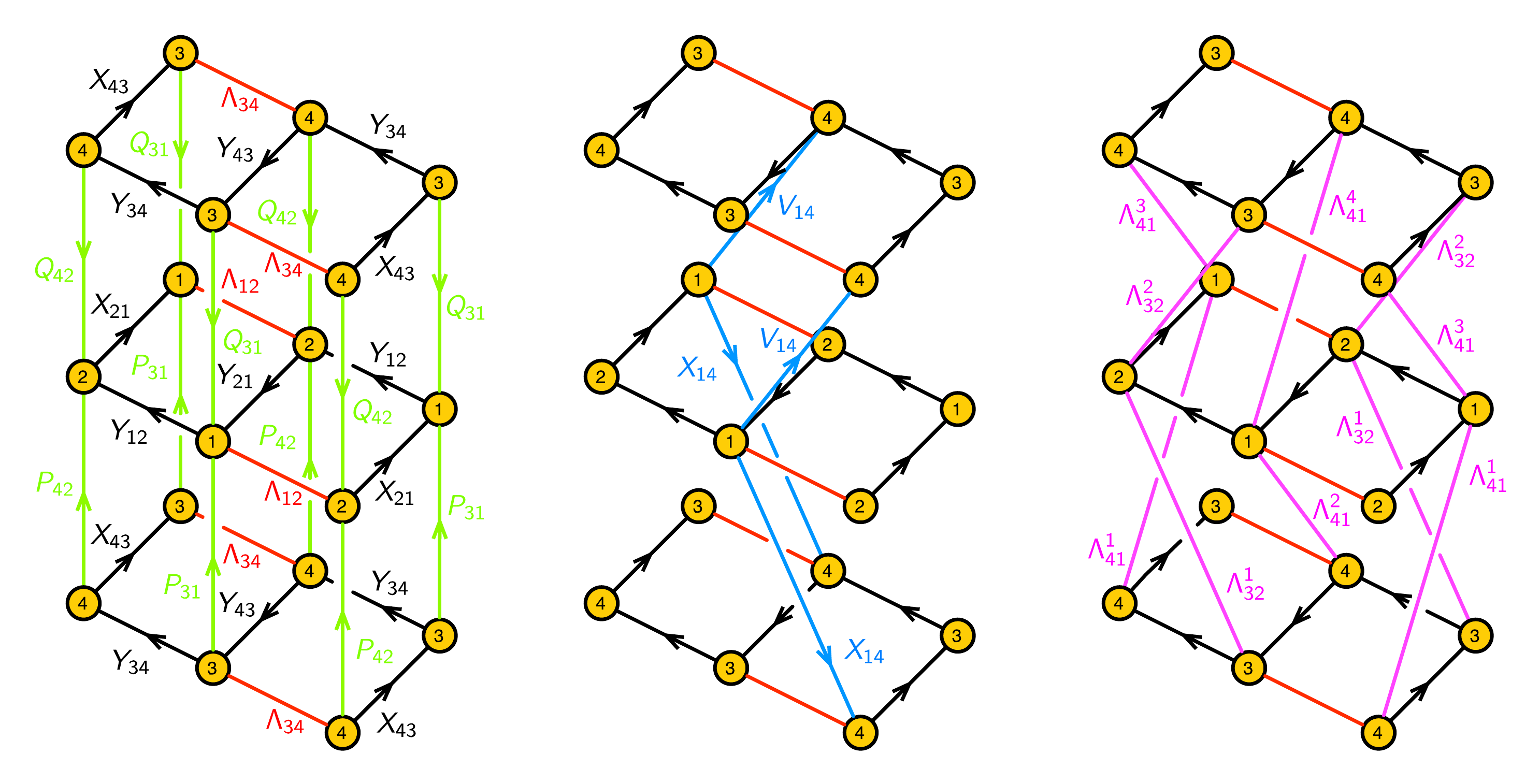}
\caption{Periodic quiver for $\mathcal{C}_{(+,-)}$. 
\label{periodic_quiver_C+-}}
 \end{center}
 \end{figure}

Let us start again from the conifold and perform orbifold reduction with  $s=(+,-)$. We refer to this theory as $\mathcal{C}_{(+,-)}$. \fref{periodic_quiver_C+-} shows its periodic quiver.

Te corresponding $J$- and $E$- terms are:
 \beq
\begin{array}{rcrcccrcc}
& & J \ \ \ \ \ \ \ \ \ \ \ \ \ \ \ \ \ & & & & E \ \ \ \ \ \ \ \ \ \ \ \ \ \ & & \\
 \Lambda_{12} : & \ \ \  & X_{21} \cdot Y_{12} \cdot Y_{21} - Y_{21} \cdot Y_{12} \cdot X_{21} & = & 0 & \ \ \ \  & V_{14} \cdot Q_{42} - X_{14} \cdot P_{42} & = & 0 \\ 
  \Lambda_{34} : & \ \ \  & X_{43} \cdot Y_{34} \cdot Y_{43} - Y_{43} \cdot Y_{34} \cdot X_{43} & = & 0 & \ \ \ \  & P_{31} \cdot X_{14} - Q_{31} \cdot V_{14} & = & 0 \\ 
 \Lambda_{41}^{1} : & \ \ \  & Y_{12} \cdot Y_{21} \cdot X_{14} - X_{14} \cdot Y_{43} \cdot Y_{34} & = & 0 & \ \ \ \  & P_{42} \cdot X_{21} - X_{43} \cdot P_{31} & = & 0 \\ 
 \Lambda_{41}^{2} : & \ \ \  & X_{14} \cdot X_{43} \cdot Y_{34} - Y_{12} \cdot X_{21} \cdot X_{14} & = & 0 & \ \ \ \  & P_{42} \cdot Y_{21} - Y_{43} \cdot P_{31} & = & 0 \\ 
 \Lambda_{41}^{3} : & \ \ \  & Y_{12} \cdot Y_{21} \cdot V_{14} - V_{14} \cdot Y_{43} \cdot Y_{34} & = & 0 & \ \ \ \  & Q_{42} \cdot X_{21} - X_{43} \cdot Q_{31} & = & 0 \\ 
 \Lambda_{41}^{4} : & \ \ \  & V_{14} \cdot X_{43} \cdot Y_{34} - Y_{12} \cdot X_{21} \cdot V_{14} & = & 0 & \ \ \ \  & Q_{42} \cdot Y_{21} - Y_{43} \cdot Q_{31} & = & 0 \\ 
 \Lambda_{32}^{1} : & \ \ \  & Y_{21} \cdot X_{14} \cdot X_{43} - X_{21} \cdot X_{14} \cdot Y_{43} & = & 0 & \ \ \ \  & P_{31} \cdot Y_{12} - Y_{34} \cdot P_{42} & = & 0 \\ 
 \Lambda_{32}^{2} : & \ \ \  & Y_{21} \cdot V_{14} \cdot X_{43} - X_{21} \cdot V_{14} \cdot Y_{43} & = & 0 & \ \ \ \  & Q_{31} \cdot Y_{12} - Y_{34} \cdot Q_{42} & = & 0 
 \end{array} 
\label{E_J_C_+-}
 \eeq

The classical mesonic moduli space for this theory corresponds to the toric diagram shown in \fref{toric_C++}, as expected. 
 
\begin{figure}[H]
\begin{center}
\includegraphics[height=6.5cm]{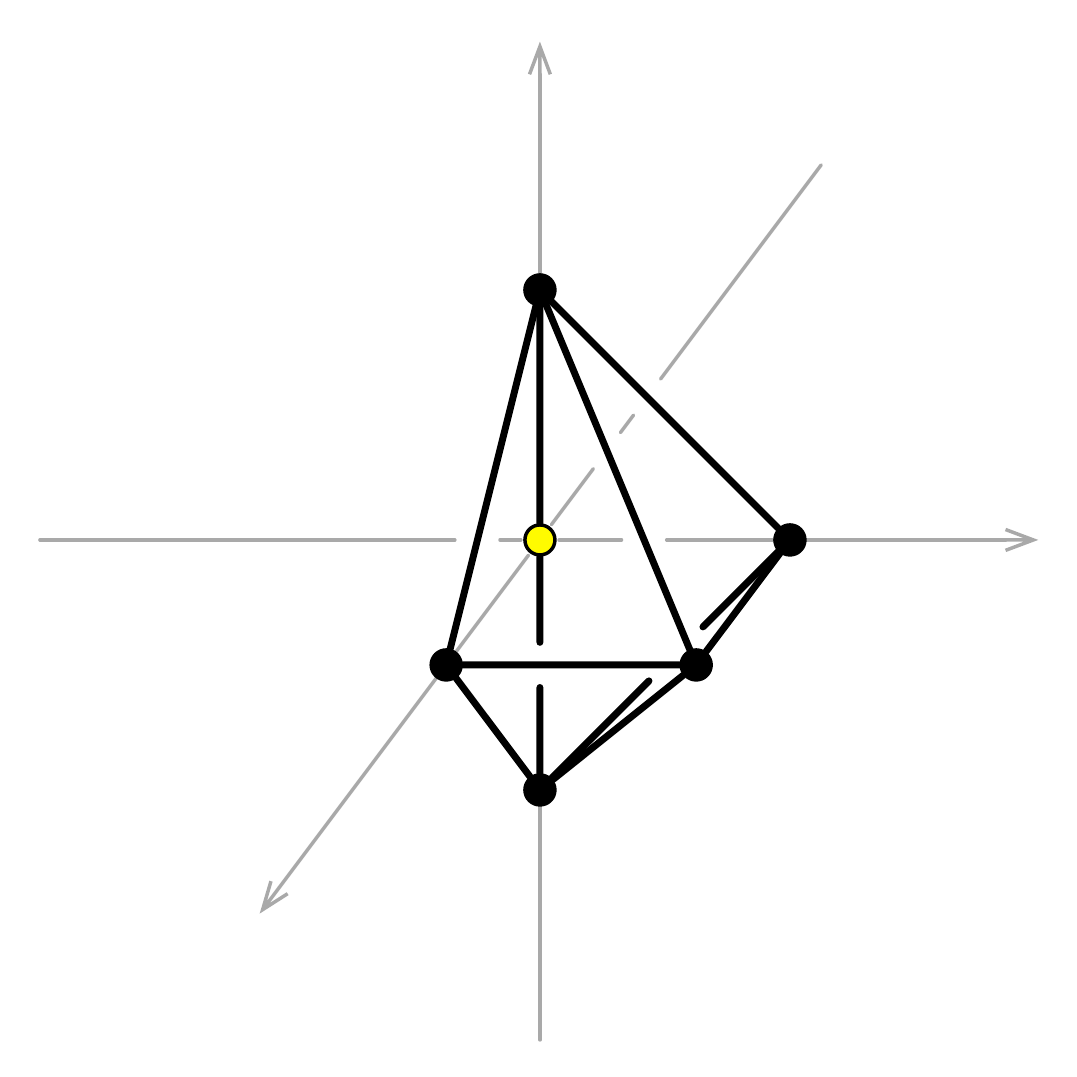}
\caption{Toric diagram for $\mathcal{C}_{(+,-)}$.}
\label{toric_C+-}
 \end{center}
 \end{figure} 

\section{Triality from Seiberg Duality}

\label{section_triality_from_Seiberg}

Orbifold reduction of Seiberg dual toric theories generically leads to different $2d$ $(0,2)$ theories associated to the same CY 4-fold. Following \cite{Franco:2016nwv,Franco:2016qxh}, we expect such theories to be related by triality (namely, either a single triality transformation or a sequence of them). This approach to triality is in the same general spirit of other constructions that derive it from Seiberg duality, such as \cite{Gadde:2015wta}.

\begin{figure}[H]
\begin{center}
\includegraphics[height=4.5cm]{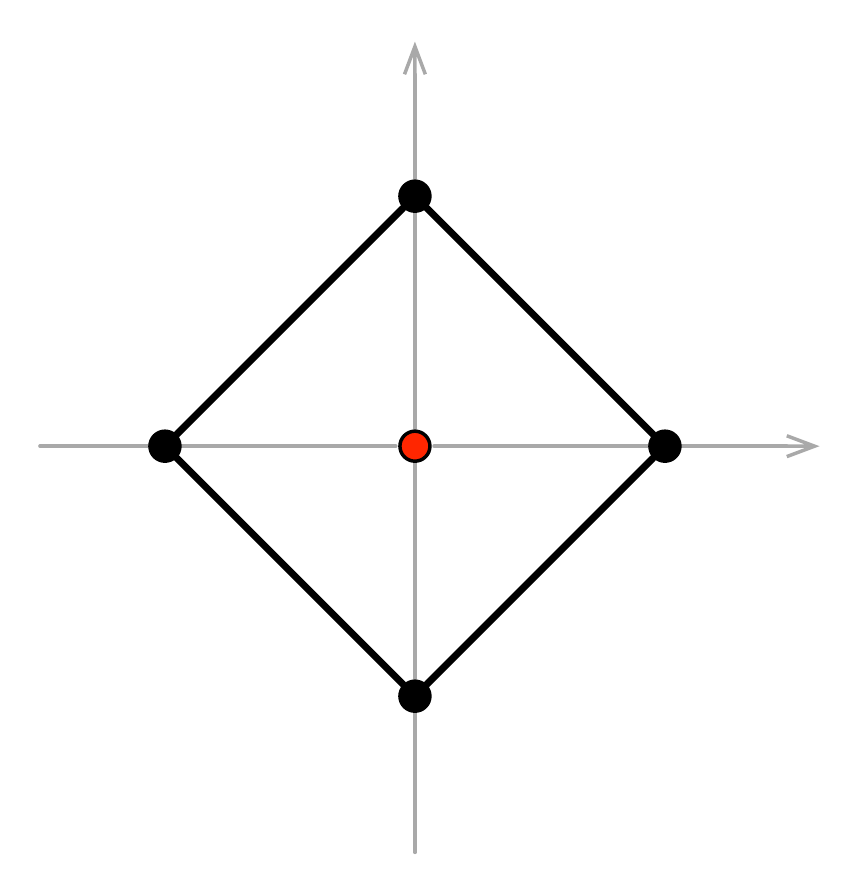}
\caption{Toric diagram for $F_0$.}
\label{toric_F0}
 \end{center}
 \end{figure}

Let us illustrate this idea with an explicit example. Consider the complex cone over $F_0$ or, for brevity, just $F_0$ from now on. Its toric diagram is shown in \fref{toric_F0}. 

\begin{figure}[ht]
\begin{center}
\includegraphics[width=9cm]{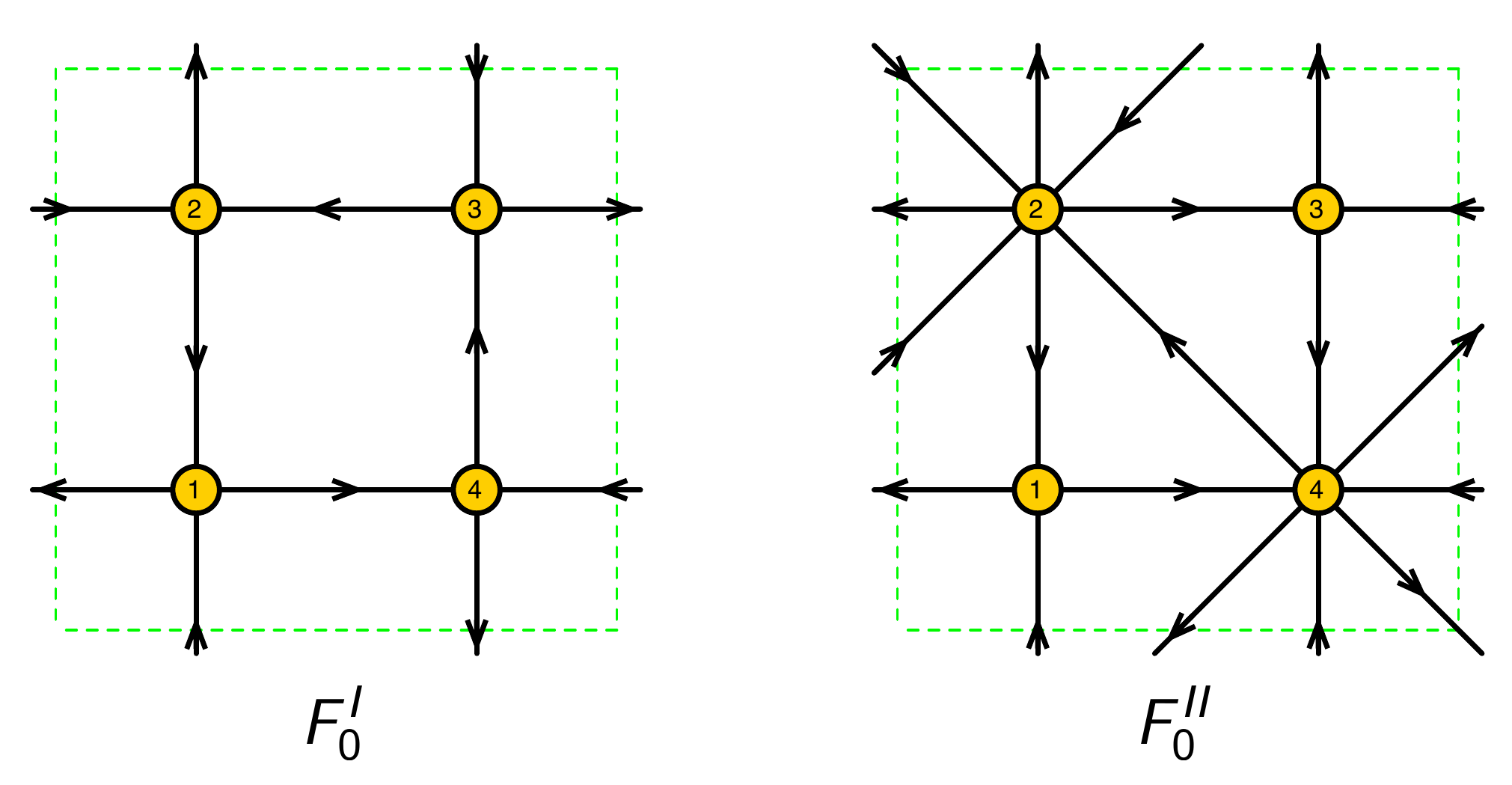}
\caption{Periodic quivers for phases I and II of $F_0$ . 
\label{dimers_F0}}
 \end{center}
 \end{figure}

There are two toric phases for $F_0$, which are related by Seiberg duality. They have been extensively studied in the literature (see e.g. \cite{Feng:2001xr,Feng:2001bn,Feng:2002zw,Franco:2005rj}). Their periodic quivers are shown in \fref{dimers_F0}.

\begin{figure}[ht]
\begin{center}
\includegraphics[width=10cm]{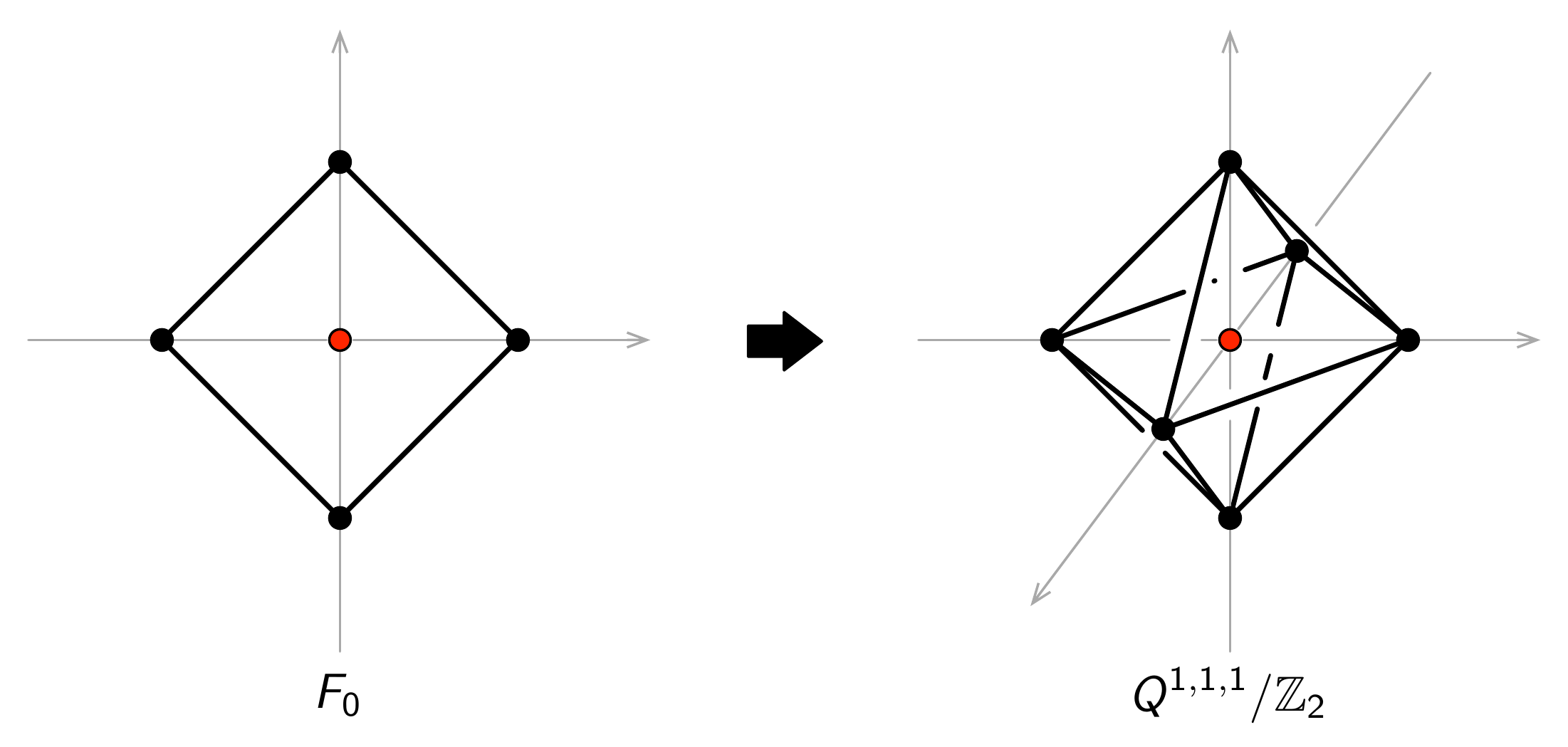}
\caption{Lift of the toric diagram from $F_0$ to  $Q^{1,1,1}/\mathbb{Z}_2$ by an $s=(+,-)$ orbifold reduction acting on the central point.
\label{toric_lift_F0_to_Q111Z2}}
 \end{center}
 \end{figure}

Starting from the two theories, we will perform orbifold reductions that lift the center point in the toric diagram with $s=(+,-)$. The toric diagram is then transformed into the one for $Q^{1,1,1}/\mathbb{Z}_2$, as shown in \fref{toric_lift_F0_to_Q111Z2}. This geometry has several toric phases connected by triality, whose study was initiated in \cite{Franco:2016nwv}.\footnote{An exhaustive classification of the toric phases of $Q^{1,1,1}/\mathbb{Z}_2$ will appear in \cite{to_appear}.} Some of these phases were analyzed using mirror symmetry in \cite{Franco:2016qxh}.

\subsection{${F_{0}^{I}}_{(+,-)}$}

\begin{figure}[H]
\begin{center}
\resizebox{1\hsize}{!}{
\includegraphics[height=3.7cm]{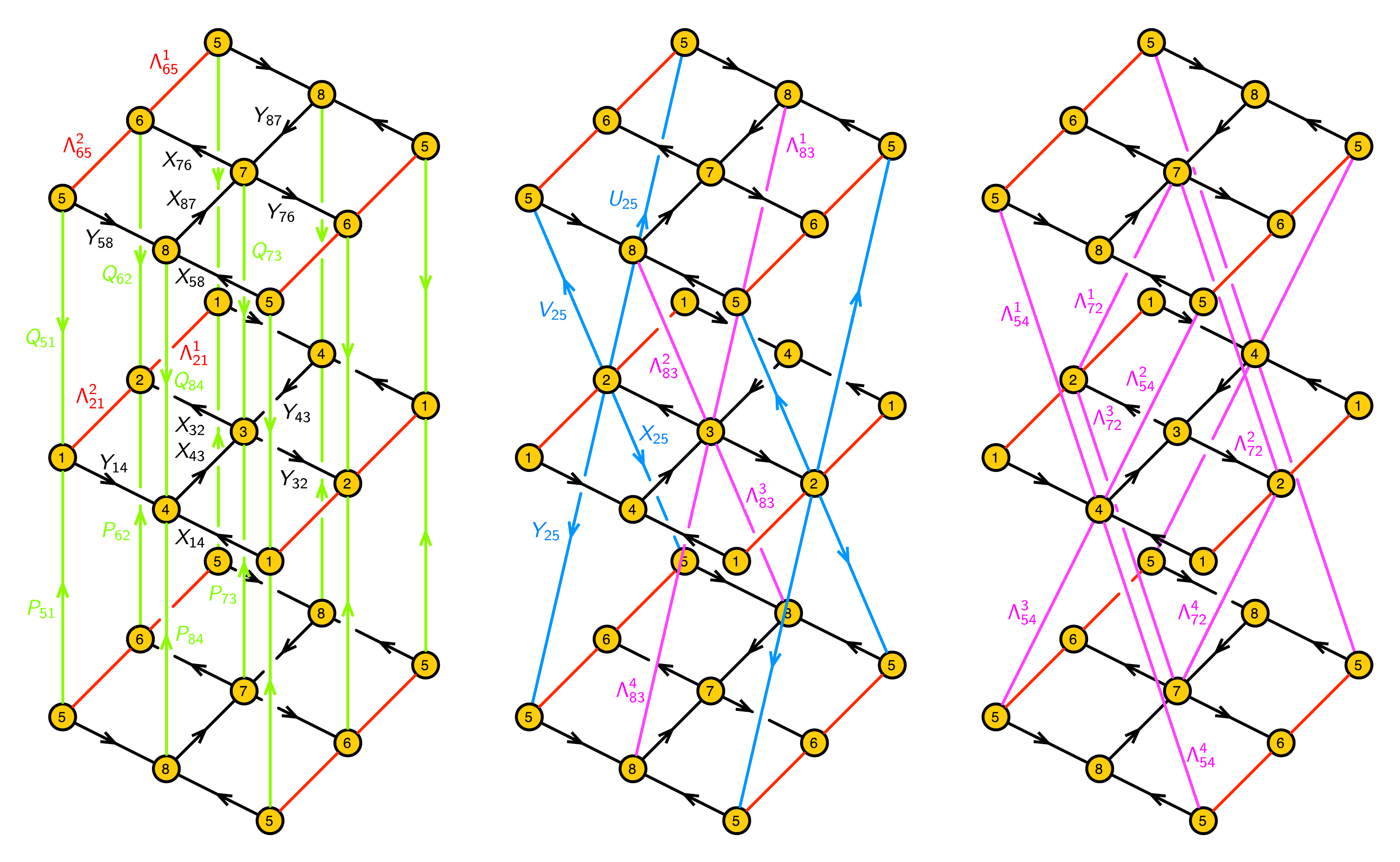}
}
\vspace{-.35cm}\caption{Periodic quiver for ${F_{0}^{I}}_{(+,-)}$.}
\label{bbf0ph1red1}
 \end{center}
 \end{figure}

Let us first consider the $s=(+,-)$ orbifold reduction of phase I of $F_0$ using a $p_0$ associated to the central point of the toric diagram. We refer to the resulting theory as ${F_{0}^{I}}_{(+,-)}$ and present its periodic quiver in \fref{bbf0ph1red1}. As explained earlier, our choice of $p_0$ can be immediately identified from this figure.

The $J$- and $E$-terms are:
{
\small
\beq
\begin{array}{rclccclcc}
& & \ \ \ \ \ \ \ \ \ \ \ \ \ \ \ \ \ \ \ \ J & & & & \ \ \ \ \ \ \ \ \ \ \ \ \ \ E & & \\
\Lambda_{65}^{1} : & \ \ \  & Y_{58} \cdot Y_{87} \cdot X_{76} - X_{58} \cdot Y_{87} \cdot Y_{76} & = & 0 & \ \ \ \  & Q_{62} \cdot U_{25} - P_{62} \cdot X_{25} & = & 0 \\
\Lambda_{72}^{1} : & \ \ \  & U_{25} \cdot Y_{58} \cdot Y_{87} - X_{87} \cdot Y_{58} \cdot V_{25} & = & 0 & \ \ \ \  & Q_{73} \cdot X_{32} - X_{76} \cdot Q_{62} & = & 0 \\
\Lambda_{83}^{1} : & \ \ \  & X_{32} \cdot U_{25} \cdot Y_{58} - X_{58} \cdot U_{25} \cdot Y_{32} & = & 0 & \ \ \ \  & Q_{84} \cdot Y_{43} - Y_{87} \cdot Q_{73} & = & 0 \\
\Lambda_{54}^{1} : & \ \ \  & Y_{43} \cdot X_{32} \cdot U_{25} - X_{43} \cdot X_{32} \cdot V_{25} & = & 0 & \ \ \ \  & Q_{51} \cdot Y_{14} - Y_{58} \cdot Q_{84} & = & 0 \\
\Lambda_{65}^{2} : & \ \ \  & X_{58} \cdot X_{87} \cdot Y_{76} - Y_{58} \cdot X_{87} \cdot X_{76} & = & 0 & \ \ \ \  & Q_{62} \cdot V_{25} - P_{62} \cdot Y_{25} & = & 0 \\
\Lambda_{72}^{2} : & \ \ \  & V_{25} \cdot X_{58} \cdot X_{87} - U_{25} \cdot X_{58} \cdot Y_{87} & = & 0 & \ \ \ \  & Q_{73} \cdot Y_{32} - Y_{76} \cdot Q_{62} & = & 0 \\
\Lambda_{83}^{2} : & \ \ \  & Y_{32} \cdot V_{25} \cdot X_{58} - X_{32} \cdot V_{25} \cdot Y_{58} & = & 0 & \ \ \ \  & Q_{84} \cdot X_{43} - X_{87} \cdot Q_{73} & = & 0 \\
\Lambda_{54}^{2} : & \ \ \  & X_{43} \cdot Y_{32} \cdot V_{25} - Y_{43} \cdot Y_{32} \cdot U_{25} & = & 0 & \ \ \ \  & Q_{51} \cdot X_{14} - X_{58} \cdot Q_{84} & = & 0 \\
\end{array} \nonumber
\eeq

\beq
\begin{array}{rclccclcc}
\Lambda_{21}^{1} : & \ \ \  & Y_{14} \cdot Y_{43} \cdot X_{32} - X_{14} \cdot Y_{43} \cdot Y_{32} & = & 0 & \ \ \ \  & X_{25} \cdot P_{51} - U_{25} \cdot Q_{51} & = & 0 \\
\Lambda_{72}^{3} : & \ \ \  & X_{25} \cdot Y_{58} \cdot Y_{87} - Y_{25} \cdot Y_{58} \cdot X_{87} & = & 0 & \ \ \ \  & P_{73} \cdot X_{32} - X_{76} \cdot P_{62} & = & 0 \\
\Lambda_{83}^{3} : & \ \ \  & X_{32} \cdot X_{25} \cdot Y_{58} - Y_{32} \cdot X_{25} \cdot X_{58} & = & 0 & \ \ \ \  & P_{84} \cdot Y_{43} - Y_{87} \cdot P_{73} & = & 0 \\
\Lambda_{54}^{3} : & \ \ \  & Y_{43} \cdot X_{32} \cdot X_{25} - X_{43} \cdot X_{32} \cdot Y_{25} & = & 0 & \ \ \ \  & P_{51} \cdot Y_{14} - Y_{58} \cdot P_{84} & = & 0 \\
\Lambda_{21}^{2} : & \ \ \  & X_{14} \cdot X_{43} \cdot Y_{32} - Y_{14} \cdot X_{43} \cdot X_{32} & = & 0 & \ \ \ \  & Y_{25} \cdot P_{51} - V_{25} \cdot Q_{51} & = & 0 \\
\Lambda_{72}^{4} : & \ \ \  & Y_{25} \cdot X_{58} \cdot X_{87} - X_{25} \cdot X_{58} \cdot Y_{87} & = & 0 & \ \ \ \  & P_{73} \cdot Y_{32} - Y_{76} \cdot P_{62} & = & 0 \\
\Lambda_{83}^{4} : & \ \ \  & Y_{32} \cdot Y_{25} \cdot X_{58} - X_{32} \cdot Y_{25} \cdot Y_{58} & = & 0 & \ \ \ \  & P_{84} \cdot X_{43} - X_{87} \cdot P_{73} & = & 0 \\
\Lambda_{54}^{4} : & \ \ \  & X_{43} \cdot Y_{32} \cdot Y_{25} - Y_{43} \cdot Y_{32} \cdot X_{25} & = & 0 & \ \ \ \  & P_{51} \cdot X_{14} - X_{58} \cdot P_{84} & = & 0 \\
\end{array}
\eeq
}

This theory corresponds to $Q^{1,1,1}/\mathbb{Z}_2$. In fact, this is phase C in the classification of \cite{Franco:2016nwv}. The ease with which orbifold reduction generates a phase for such a complicated geometry is truly remarkable.

\subsection{${F_{0}^{II}}_{(+,-)}$}

\begin{figure}[H]
\begin{center}
\resizebox{1\hsize}{!}{
\includegraphics[height=3.7cm]{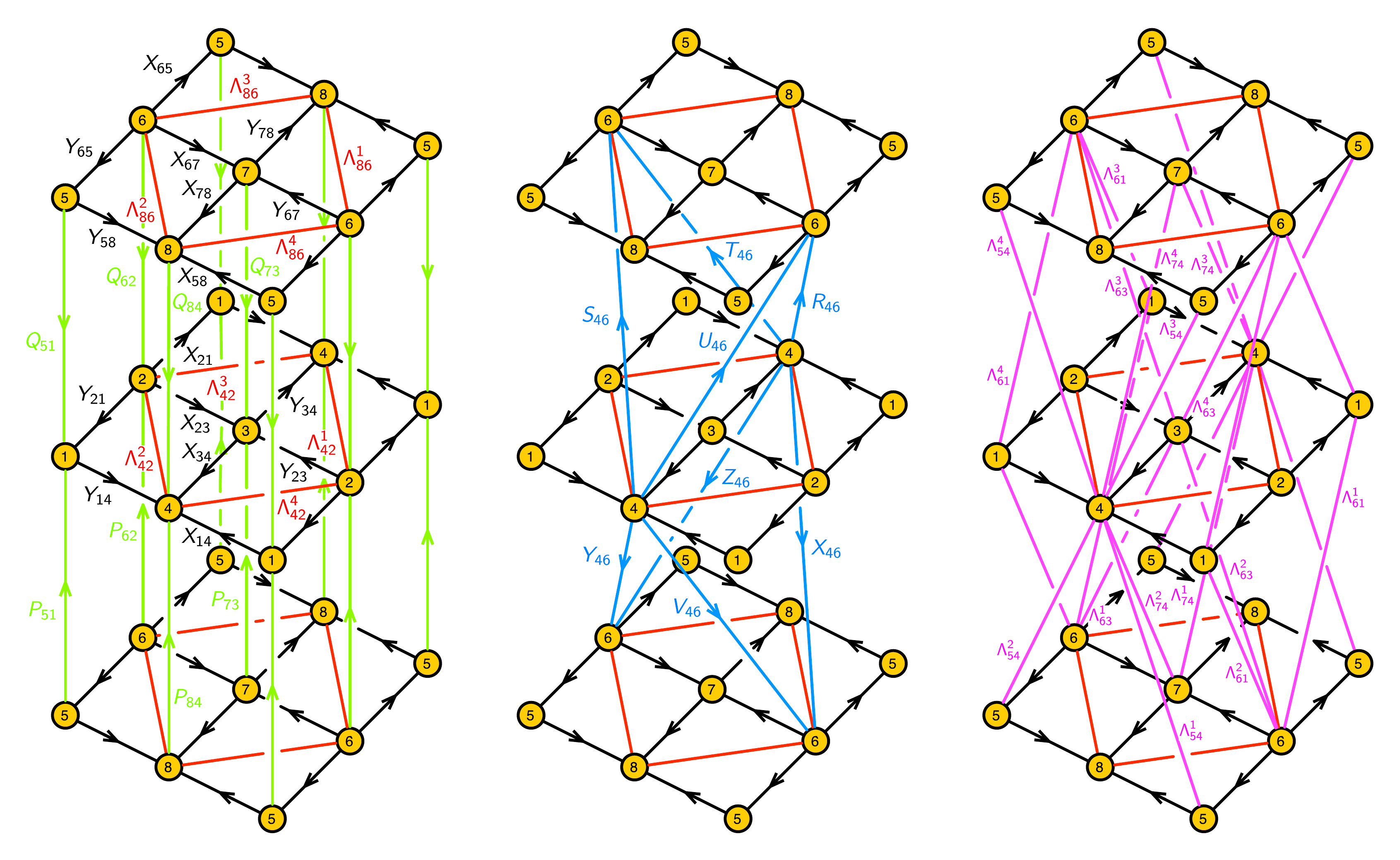}
}
\vspace{-.35cm}\caption{Periodic quiver for ${F_{0}^{II}}_{(+,-)}$.}
\label{bbf0ph2red1}
 \end{center}
 \end{figure}

Next, let us construct a second theory for $Q^{1,1,1}/\mathbb{Z}_2$, by performing the orbifold reduction of phase II of $F_0$ with $s=(+,-)$ and a $p_0$ associated to the central point in the toric diagram. We call this theory ${F_{0}^{II}}_{(+,-)}$ and present its periodic quiver in \fref{bbf0ph2red1}.

The $J$- and $E$-terms are:
{
\small
\beq
\begin{array}{rclccclcc}
& & \ \ \ \ \ \ \ \ \ \ \ \ \ \ \ J & & & & \ \ \ \ \ \ \ \ \ \ \ \ \ \ E & & \\
 \Lambda_{61}^{1} : & \ \ \  & X_{14} \cdot X_{46} - Y_{14} \cdot Z_{46} & = & 0 & \ \ \ \  & X_{65} \cdot P_{51} - P_{62} \cdot X_{21} & = & 0 \\ 
 \Lambda_{54}^{1} : & \ \ \  & X_{46} \cdot X_{65} - V_{46} \cdot Y_{65} & = & 0 & \ \ \ \  & X_{58} \cdot P_{84} - P_{51} \cdot X_{14} & = & 0 \\ 
 \Lambda_{86}^{1} : & \ \ \  & X_{65} \cdot X_{58} - Y_{67} \cdot Y_{78} & = & 0 & \ \ \ \  & P_{84} \cdot X_{46} - Q_{84} \cdot R_{46} & = & 0 \\  
 \Lambda_{61}^{2} : & \ \ \  & Y_{14} \cdot Y_{46} - X_{14} \cdot V_{46} & = & 0 & \ \ \ \  & Y_{65} \cdot P_{51} - P_{62} \cdot Y_{21} & = & 0 \\ 
 \Lambda_{54}^{2} : & \ \ \  & Y_{46} \cdot Y_{65} - Z_{46} \cdot X_{65} & = & 0 & \ \ \ \  & Y_{58} \cdot P_{84} - P_{51} \cdot Y_{14} & = & 0 \\ 
 \Lambda_{86}^{2} : & \ \ \  & Y_{65} \cdot Y_{58} - X_{67} \cdot X_{78} & = & 0 & \ \ \ \  & P_{84} \cdot Y_{46} - Q_{84} \cdot S_{46} & = & 0 \\ 
  \Lambda_{63}^{1} : & \ \ \  & Y_{34} \cdot Z_{46} - X_{34} \cdot Y_{46} & = & 0 & \ \ \ \  & X_{67} \cdot P_{73} - P_{62} \cdot X_{23} & = & 0 \\ 
 \Lambda_{74}^{1} : & \ \ \  & Z_{46} \cdot X_{67} - X_{46} \cdot Y_{67} & = & 0 & \ \ \ \  & Y_{78} \cdot P_{84} - P_{73} \cdot Y_{34} & = & 0 \\ 
  \Lambda_{86}^{3} : & \ \ \  & X_{67} \cdot Y_{78} - X_{65} \cdot Y_{58} & = & 0 & \ \ \ \  & P_{84} \cdot Z_{46} - Q_{84} \cdot T_{46} & = & 0 \\ 
 \Lambda_{63}^{2} : & \ \ \  & X_{34} \cdot V_{46} - Y_{34} \cdot X_{46} & = & 0 & \ \ \ \  & Y_{67} \cdot P_{73} - P_{62} \cdot Y_{23} & = & 0 \\ 
  \Lambda_{74}^{2} : & \ \ \  & V_{46} \cdot Y_{67} - Y_{46} \cdot X_{67} & = & 0 & \ \ \ \  & X_{78} \cdot P_{84} - P_{73} \cdot X_{34} & = & 0 \\ 
 \Lambda_{86}^{4} : & \ \ \  & Y_{67} \cdot X_{78} - Y_{65} \cdot X_{58} & = & 0 & \ \ \ \  & P_{84} \cdot V_{46} - Q_{84} \cdot U_{46} & = & 0 \\ 
  \Lambda_{61}^{3} : & \ \ \  & X_{14} \cdot R_{46} - Y_{14} \cdot T_{46} & = & 0 & \ \ \ \  & X_{65} \cdot Q_{51} - Q_{62} \cdot X_{21} & = & 0 \\ 
 \Lambda_{54}^{3} : & \ \ \  & R_{46} \cdot X_{65} - U_{46} \cdot Y_{65} & = & 0 & \ \ \ \  & X_{58} \cdot Q_{84} - Q_{51} \cdot X_{14} & = & 0 \\ 
  \Lambda_{42}^{1} : & \ \ \  & X_{21} \cdot X_{14} - Y_{23} \cdot Y_{34} & = & 0 & \ \ \ \  & R_{46} \cdot Q_{62} - X_{46} \cdot P_{62} & = & 0 \\ 
 \Lambda_{61}^{4} : & \ \ \  & Y_{14} \cdot S_{46} - X_{14} \cdot U_{46} & = & 0 & \ \ \ \  & Y_{65} \cdot Q_{51} - Q_{62} \cdot Y_{21} & = & 0 \\ 
  \Lambda_{54}^{4} : & \ \ \  & S_{46} \cdot Y_{65} - T_{46} \cdot X_{65} & = & 0 & \ \ \ \  & Y_{58} \cdot Q_{84} - Q_{51} \cdot Y_{14} & = & 0 \\ 
 \Lambda_{42}^{2} : & \ \ \  & Y_{21} \cdot Y_{14} - X_{23} \cdot X_{34} & = & 0 & \ \ \ \  & S_{46} \cdot Q_{62} - Y_{46} \cdot P_{62} & = & 0 \\ 
 \Lambda_{63}^{3} : & \ \ \  & Y_{34} \cdot T_{46} - X_{34} \cdot S_{46} & = & 0 & \ \ \ \  & X_{67} \cdot Q_{73} - Q_{62} \cdot X_{23} & = & 0 \\ 
 \Lambda_{74}^{3} : & \ \ \  & T_{46} \cdot X_{67} - R_{46} \cdot Y_{67} & = & 0 & \ \ \ \  & Y_{78} \cdot Q_{84} - Q_{73} \cdot Y_{34} & = & 0 \\
 \Lambda_{42}^{3} : & \ \ \  & X_{23} \cdot Y_{34} - X_{21} \cdot Y_{14} & = & 0 & \ \ \ \  & T_{46} \cdot Q_{62} - Z_{46} \cdot P_{62} & = & 0 \\ 
 \Lambda_{63}^{4} : & \ \ \  & X_{34} \cdot U_{46} - Y_{34} \cdot R_{46} & = & 0 & \ \ \ \  & Y_{67} \cdot Q_{73} - Q_{62} \cdot Y_{23} & = & 0 \\ 
 \Lambda_{74}^{4} : & \ \ \  & U_{46} \cdot Y_{67} - S_{46} \cdot X_{67} & = & 0 & \ \ \ \  & X_{78} \cdot Q_{84} - Q_{73} \cdot X_{34} & = & 0 \\ 
 \Lambda_{42}^{4} : & \ \ \  & Y_{23} \cdot X_{34} - Y_{21} \cdot X_{14} & = & 0 & \ \ \ \  & U_{46} \cdot Q_{62} - V_{46} \cdot P_{62} & = & 0 \\
\end{array}
\eeq
}

This is a new theory that has not appeared in the existing partial survey of phases of $Q^{1,1,1}/\mathbb{Z}_2$ of \cite{Franco:2016nwv}. It is possible to show that it is indeed related by triality to the other phases. In particular, it can be obtained from ${F_{0}^{I}}_{(+,-)}$ (phase C) by consecutive triality transformations on nodes 1 and 4 and relabeling of nodes.\footnote{We are thankful to Azeem Hasan for his help establishing this fact and for his collaboration in the related classification of $Q^{1,1,1}/\mathbb{Z}_2$ phases that will appear in \cite{to_appear}.}

The two theories considered in this section demonstrate one of the salient features of orbifold reduction: the simplicity with which it generates $2d$ $(0,2)$ gauge theories for relatively complicated CY 4-folds. They also show that, in general, not all toric phases of a CY$_4$ can be derived by orbifold reduction. In particular, only phases whose field contents are related to $4d$ parents by \eref{2d_field_content_from_4d} can be constructed this way.

\section{Triality from Combinatorics in Multi-Layered Orbifold Reduction}

\label{section_triality_combinatorics}

Things become very interesting when the number of layers $k$ is increased. Whenever we have a pair of orbifold reduced theories obtained by lifting the same point in the toric diagram and whose sign vectors are related by permutations but not by an overall sign flip, we expect them to be different gauge theories that correspond to the same CY$_4$, and hence to be related by triality. This phenomenon first arises for $k=4$, for which $s=(+,+,-,-)$ and $s=(+,-,+,-)$ produce different gauge theories. 

\begin{figure}[H]
\begin{center}
\includegraphics[width=6.5cm]{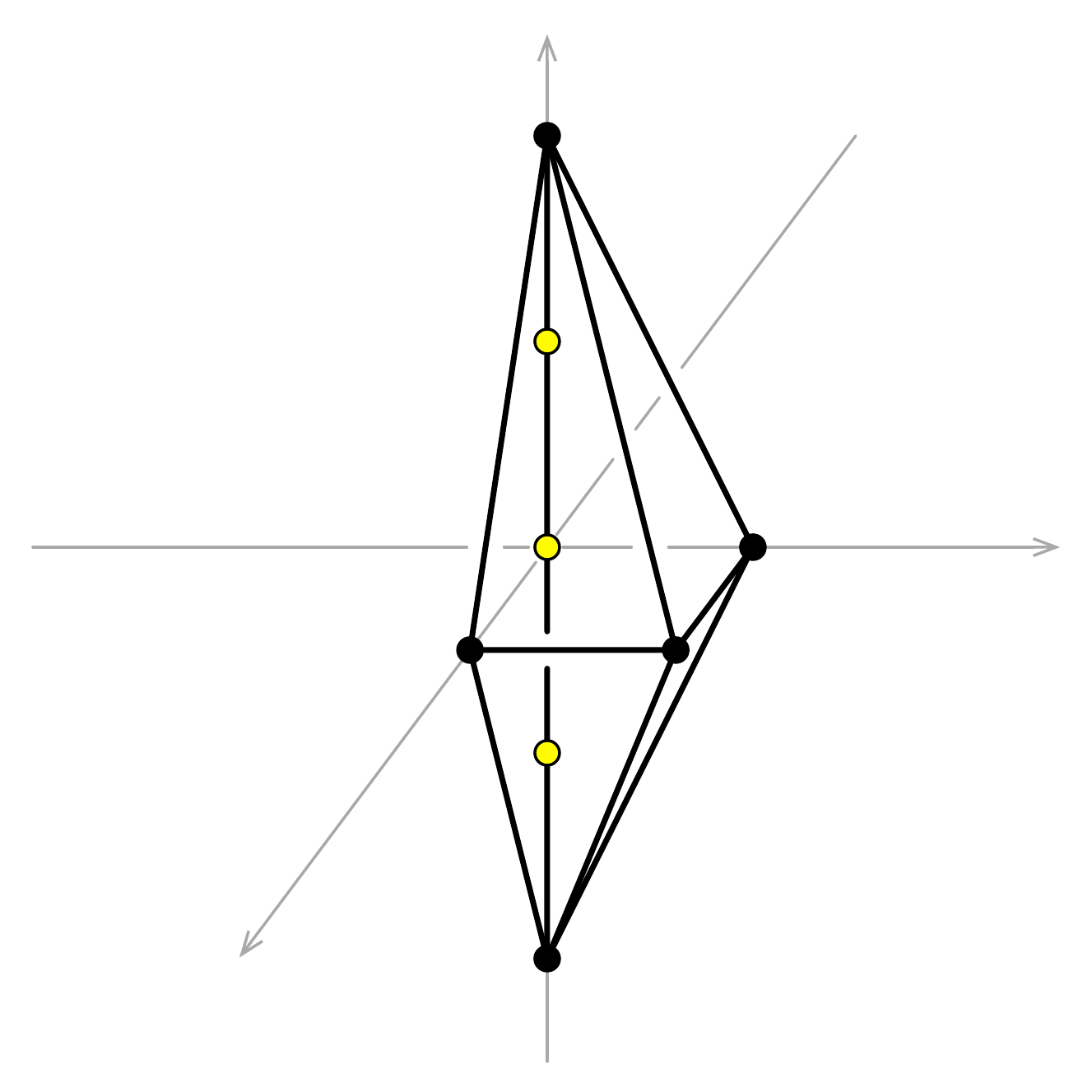}
\caption{Toric diagram for both the $\mathcal{C}_{(+,+,-,-)}$ and $\mathcal{C}_{(+,-,+,-)}$ theories.
\label{toricconifoldc}}
 \end{center}
 \end{figure}

Below we consider two explicit examples based on the conifold: $\mathcal{C}_{(+,+,-,-)}$ and $\mathcal{C}_{(+,-,+,-)}$. We have explicitly verified that classical mesonic moduli spaces for both of them correspond to the toric diagram shown in \fref{toricconifoldc}. 

\subsection{$\mathcal{C}_{(+,+,-,-)}$}

\begin{figure}[ht]
\begin{center}
\includegraphics[height=9cm]{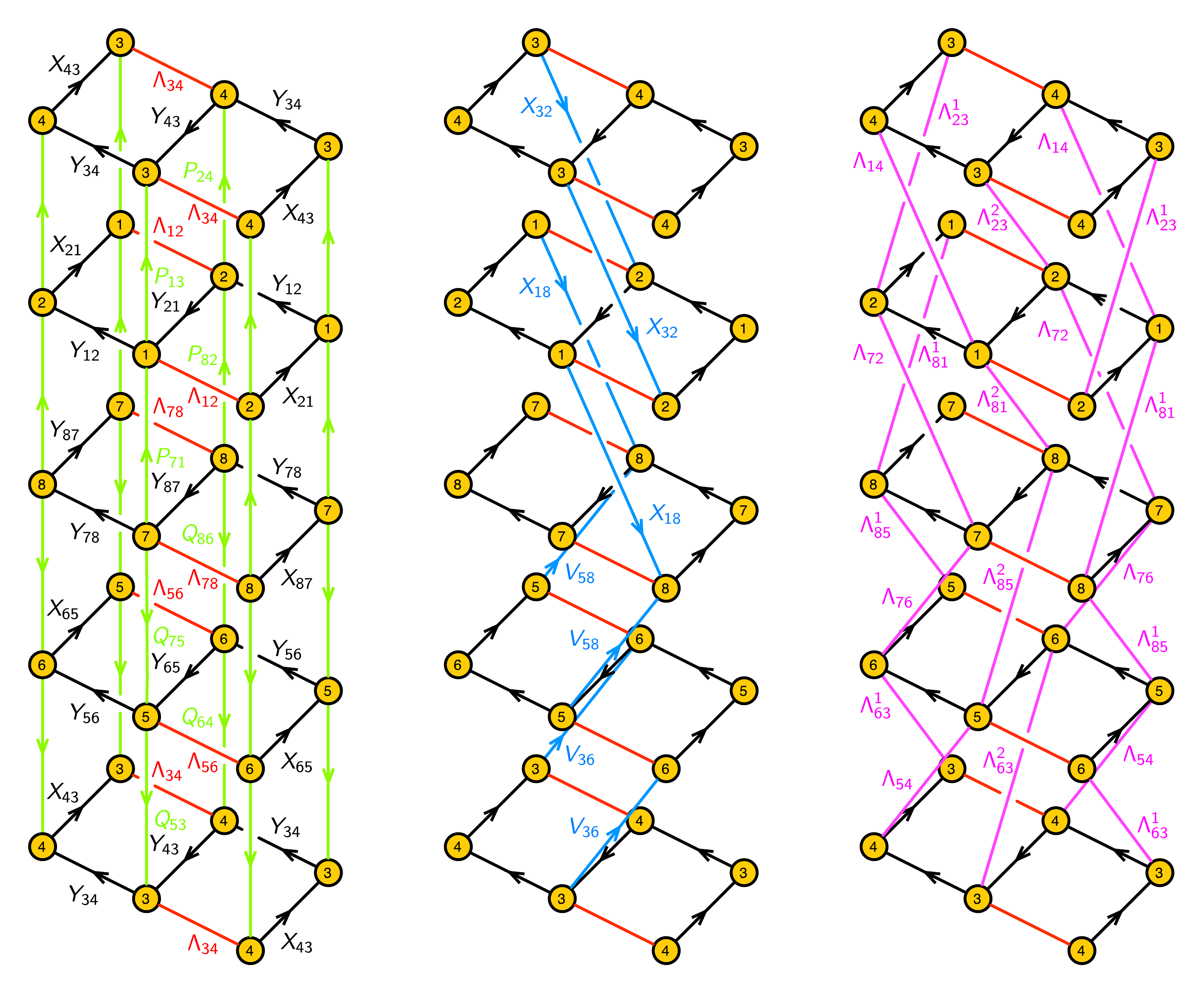}
\caption{Periodic quiver for $\mathcal{C}_{(+,+,-,-)}$. 
\label{quiver_C++--}}
 \end{center}
 \end{figure}

The periodic quiver for this theory is shown in \fref{quiver_C++--}. The $J$- and $E$-terms are:
{\small
\beq
\begin{array}{rclccclcc}
& & \ \ \ \ \ \ \ \ \ \ \ \ \ \ \ \ \ \ \ \ J & & & & \ \ \ \ \ \ \ \ \ \ \ \ \ \ E & & \\
\Lambda_{12} : & \ \ \  & X_{21} \cdot Y_{12} \cdot Y_{21} - Y_{21} \cdot Y_{12} \cdot X_{21} & = & 0 & \ \ \ \  & P_{13} \cdot X_{32} - X_{18} \cdot P_{82} & = & 0 \\
\Lambda_{81}^{1} : & \ \ \  & Y_{12} \cdot Y_{21} \cdot X_{18} - X_{18} \cdot Y_{87} \cdot Y_{78} & = & 0 & \ \ \ \  & P_{82} \cdot X_{21} - X_{87} \cdot P_{71} & = & 0 \\
\Lambda_{72} : & \ \ \  & Y_{21} \cdot X_{18} \cdot X_{87} - X_{21} \cdot X_{18} \cdot Y_{87} & = & 0 & \ \ \ \  & P_{71} \cdot Y_{12} - Y_{78} \cdot P_{82} & = & 0 \\
\Lambda_{81}^{2} : & \ \ \  & X_{18} \cdot X_{87} \cdot Y_{78} - Y_{12} \cdot X_{21} \cdot X_{18} & = & 0 & \ \ \ \  & P_{82} \cdot Y_{21} - Y_{87} \cdot P_{71} & = & 0 \\
\Lambda_{34} : & \ \ \  & X_{43} \cdot Y_{34} \cdot Y_{43} - Y_{43} \cdot Y_{34} \cdot X_{43} & = & 0 & \ \ \ \  & V_{36} \cdot Q_{64} - X_{32} \cdot P_{24} & = & 0 \\
\Lambda_{23}^{1} : & \ \ \  & Y_{34} \cdot Y_{43} \cdot X_{32} - X_{32} \cdot Y_{21} \cdot Y_{12} & = & 0 & \ \ \ \  & P_{24} \cdot X_{43} - X_{21} \cdot P_{13} & = & 0 \\
 \end{array} \nonumber
 \eeq

\beq
\begin{array}{rclccclcc}
\Lambda_{14} : & \ \ \  & Y_{43} \cdot X_{32} \cdot X_{21} - X_{43} \cdot X_{32} \cdot Y_{21} & = & 0 & \ \ \ \  & P_{13} \cdot Y_{34} - Y_{12} \cdot P_{24} & = & 0 \\
\Lambda_{23}^{2} : & \ \ \  & X_{32} \cdot X_{21} \cdot Y_{12} - Y_{34} \cdot X_{43} \cdot X_{32} & = & 0 & \ \ \ \  & P_{24} \cdot Y_{43} - Y_{21} \cdot P_{13} & = & 0 \\
\Lambda_{56} : & \ \ \  & X_{65} \cdot Y_{56} \cdot Y_{65} - Y_{65} \cdot Y_{56} \cdot X_{65} & = & 0 & \ \ \ \  & Q_{53} \cdot V_{36} - V_{58} \cdot Q_{86} & = & 0 \\
\Lambda_{63}^{1} : & \ \ \  & Y_{34} \cdot Y_{43} \cdot V_{36} - V_{36} \cdot Y_{65} \cdot Y_{56} & = & 0 & \ \ \ \  & Q_{64} \cdot X_{43} - X_{65} \cdot Q_{53} & = & 0 \\
\Lambda_{54} : & \ \ \  & Y_{43} \cdot V_{36} \cdot X_{65} - X_{43} \cdot V_{36} \cdot Y_{65} & = & 0 & \ \ \ \  & Q_{53} \cdot Y_{34} - Y_{56} \cdot Q_{64} & = & 0 \\
\Lambda_{63}^{2} : & \ \ \  & V_{36} \cdot X_{65} \cdot Y_{56} - Y_{34} \cdot X_{43} \cdot V_{36} & = & 0 & \ \ \ \  & Q_{64} \cdot Y_{43} - Y_{65} \cdot Q_{53} & = & 0 \\
\Lambda_{78} : & \ \ \  & X_{87} \cdot Y_{78} \cdot Y_{87} - Y_{87} \cdot Y_{78} \cdot X_{87} & = & 0 & \ \ \ \  & P_{71} \cdot X_{18} - Q_{75} \cdot V_{58} & = & 0 \\
\Lambda_{85}^{1} : & \ \ \  & Y_{56} \cdot Y_{65} \cdot V_{58} - V_{58} \cdot Y_{87} \cdot Y_{78} & = & 0 & \ \ \ \  & Q_{86} \cdot X_{65} - X_{87} \cdot Q_{75} & = & 0 \\
\Lambda_{76} : & \ \ \  & Y_{65} \cdot V_{58} \cdot X_{87} - X_{65} \cdot V_{58} \cdot Y_{87} & = & 0 & \ \ \ \  & Q_{75} \cdot Y_{56} - Y_{78} \cdot Q_{86} & = & 0 \\
\Lambda_{85}^{2} : & \ \ \  & V_{58} \cdot X_{87} \cdot Y_{78} - Y_{56} \cdot X_{65} \cdot V_{58} & = & 0 & \ \ \ \  & Q_{86} \cdot Y_{65} - Y_{87} \cdot Q_{75} & = & 0 \\
 \end{array}
 \eeq
 }

\subsection{$\mathcal{C}_{(+,-,+,-)}$}

\begin{figure}[H]
\begin{center}
\includegraphics[height=8cm]{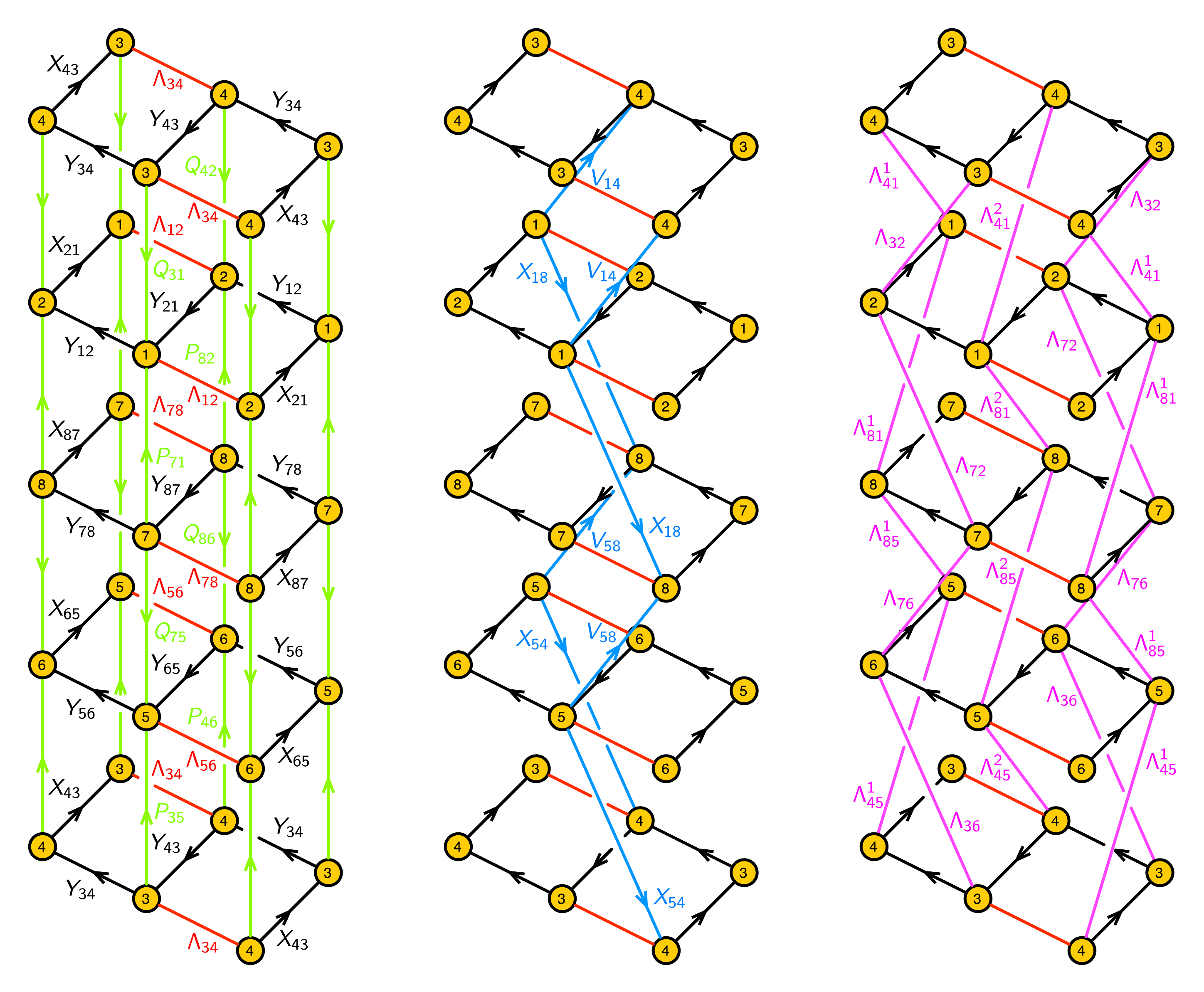}
\caption{Periodic quiver for $\mathcal{C}_{(+,-,+,-)}$.
\label{periodic_quiver_C+-+-}}
 \end{center}
 \end{figure}

The periodic quiver for this theory is shown in \fref{periodic_quiver_C+-+-}.

The $J$- and $E$-terms are:
{\small
\beq
\begin{array}{rclccclcc}
& & \ \ \ \ \ \ \ \ \ \ \ \ \ \ \ \ \ \ \ \ J & & & & \ \ \ \ \ \ \ \ \ \ \ \ \ \ E & & \\
\Lambda_{12} : & \ \ \  & X_{21} \cdot Y_{12} \cdot Y_{21}  - Y_{21} \cdot Y_{12} \cdot X_{21} & = & 0 & \ \ \ \  & X_{18} \cdot P_{82}  - V_{14} \cdot Q_{42} & = & 0 \\
\Lambda_{81}^{1} : & \ \ \  & Y_{12} \cdot Y_{21} \cdot X_{18}  - X_{18} \cdot Y_{87} \cdot Y_{78} & = & 0 & \ \ \ \  & P_{82} \cdot X_{21}  - X_{87} \cdot P_{71} & = & 0 \\
\Lambda_{72} : & \ \ \  & Y_{21} \cdot X_{18} \cdot X_{87}  - X_{21} \cdot X_{18} \cdot Y_{87} & = & 0 & \ \ \ \  & P_{71} \cdot Y_{12}  - Y_{78} \cdot P_{82} & = & 0 \\
\Lambda_{81}^{2} : & \ \ \  & X_{18} \cdot X_{87} \cdot Y_{78}  - Y_{12} \cdot X_{21} \cdot X_{18} & = & 0 & \ \ \ \  & P_{82} \cdot Y_{21}  - Y_{87} \cdot P_{71} & = & 0 \\
\Lambda_{34} : & \ \ \  & X_{43} \cdot Y_{34} \cdot Y_{43}  - Y_{43} \cdot Y_{34} \cdot X_{43} & = & 0 & \ \ \ \  & Q_{31} \cdot V_{14}  - P_{35} \cdot X_{54} & = & 0 \\
\Lambda_{41}^{1} : & \ \ \  & Y_{12} \cdot Y_{21} \cdot V_{14}  - V_{14} \cdot Y_{43} \cdot Y_{34} & = & 0 & \ \ \ \  & Q_{42} \cdot X_{21}  - X_{43} \cdot Q_{31} & = & 0 \\
\Lambda_{32} : & \ \ \  & Y_{21} \cdot V_{14} \cdot X_{43}  - X_{21} \cdot V_{14} \cdot Y_{43} & = & 0 & \ \ \ \  & Q_{31} \cdot Y_{12}  - Y_{34} \cdot Q_{42} & = & 0 \\
\Lambda_{41}^{2} : & \ \ \  & V_{14} \cdot X_{43} \cdot Y_{34}  - Y_{12} \cdot X_{21} \cdot V_{14} & = & 0 & \ \ \ \  & Q_{42} \cdot Y_{21}  - Y_{43} \cdot Q_{31} & = & 0 \\
\Lambda_{56} : & \ \ \  & X_{65} \cdot Y_{56} \cdot Y_{65}  - Y_{65} \cdot Y_{56} \cdot X_{65} & = & 0 & \ \ \ \  & V_{58} \cdot Q_{86}  - X_{54} \cdot P_{46} & = & 0 \\
\Lambda_{45}^{1} : & \ \ \  & Y_{56} \cdot Y_{65} \cdot X_{54}  - X_{54} \cdot Y_{43} \cdot Y_{34} & = & 0 & \ \ \ \  & P_{46} \cdot X_{65}  - X_{43} \cdot P_{35} & = & 0 \\
\Lambda_{36} : & \ \ \  & Y_{65} \cdot X_{54} \cdot X_{43}  - X_{65} \cdot X_{54} \cdot Y_{43} & = & 0 & \ \ \ \  & P_{35} \cdot Y_{56}  - Y_{34} \cdot P_{46} & = & 0 \\
\Lambda_{45}^{2} : & \ \ \  & X_{54} \cdot X_{43} \cdot Y_{34}  - Y_{56} \cdot X_{65} \cdot X_{54} & = & 0 & \ \ \ \  & P_{46} \cdot Y_{65}  - Y_{43} \cdot P_{35} & = & 0 \\
\Lambda_{78} : & \ \ \  & X_{87} \cdot Y_{78} \cdot Y_{87}  - Y_{87} \cdot Y_{78} \cdot X_{87} & = & 0 & \ \ \ \  & Q_{75} \cdot V_{58}  - P_{71} \cdot X_{18} & = & 0 \\
\Lambda_{85}^{1} : & \ \ \  & Y_{56} \cdot Y_{65} \cdot V_{58}  - V_{58} \cdot Y_{87} \cdot Y_{78} & = & 0 & \ \ \ \  & Q_{86} \cdot X_{65}  - X_{87} \cdot Q_{75} & = & 0 \\
\Lambda_{76} : & \ \ \  & Y_{65} \cdot V_{58} \cdot X_{87}  - X_{65} \cdot V_{58} \cdot Y_{87} & = & 0 & \ \ \ \  & Q_{75} \cdot Y_{56}  - Y_{78} \cdot Q_{86} & = & 0 \\
\Lambda_{85}^{2} : & \ \ \  & V_{58} \cdot X_{87} \cdot Y_{78}  - Y_{56} \cdot X_{65} \cdot V_{58} & = & 0 & \ \ \ \  & Q_{86} \cdot Y_{65}  - Y_{87} \cdot Q_{75} & = & 0 \\
 \end{array}
 \eeq
 }

It is possible to show that the two theories discussed above are related by triality, as expected. Explicitly: starting from $\mathcal{C}_{(+,+,-,-)}$, performing a triality transformation on node 7, relabeling nodes according to $(1,2,3,4,5,6,7,8) \to (2,1,4,3,6,5,7,8)$ and charge conjugating all fields, we obtain precisely the $\mathcal{C}_{(+,-,+,-)}$ theory.

The field content and types of interactions terms in the two theories are very similar. The action of the permutation of the sign vectors as a rearrangement of layers along the vertical direction of the periodic quiver is reminiscent of the motion of impurities in $Y^{p,q}$ and $L^{a,b,a}$ quivers due to Seiberg duality \cite{Benvenuti:2004dy,Herzog:2004tr,Benvenuti:2004wx,Franco:2005sm}. It would be interesting to explore this connection in further detail.

\section{2d (0,2) Klebanov-Witten Deformations}

\label{section_KW}

In their seminal paper \cite{Klebanov:1998hh}, Klebanov and Witten (KW) introduced a mass deformation that connects the $4d$ $\mathcal{N}=2$ gauge theory on D3-branes probing $\mathbb{C}^2/\mathbb{Z}_2 \times \mathbb{C}$ to the $\mathcal{N}=1$ gauge theory associated to the conifold. This type of deformation has been generalized to a wide class of $4d$ theories on D3-branes probing toric CY$_3$ singularities and studied in detail in terms of brane tilings \cite{Bianchi:2014qma}. The new examples include starting points with only $\mathcal{N}=1$ SUSY. 

It is natural to ask whether $2d$ analogues of KW deformations exist: namely mass deformations connecting the gauge theories on D1-branes probing two different CY$_4$ geometries.\footnote{We are indebted to Igor Klebanov for raising this question.} First, it is easy to verify that the dimensional reductions of $4d$ KW-type deformations work. Such deformations, connect pairs of theories with at least $(2,2)$ SUSY. A more interesting question is whether $(0,2)$ KW deformations exist. Below we present an explicit example in which both the initial and final theories have $(0,2)$ SUSY. It would certainly be interesting to carry out a systematic investigation of $(0,2)$ KW deformations. We leave this question for future work.

\subsection{A Deformation from $\mathcal{C}_{(+,-)}$ to $Q^{1,1,1}$}

\begin{figure}[ht]
\begin{center}
\includegraphics[width=12cm]{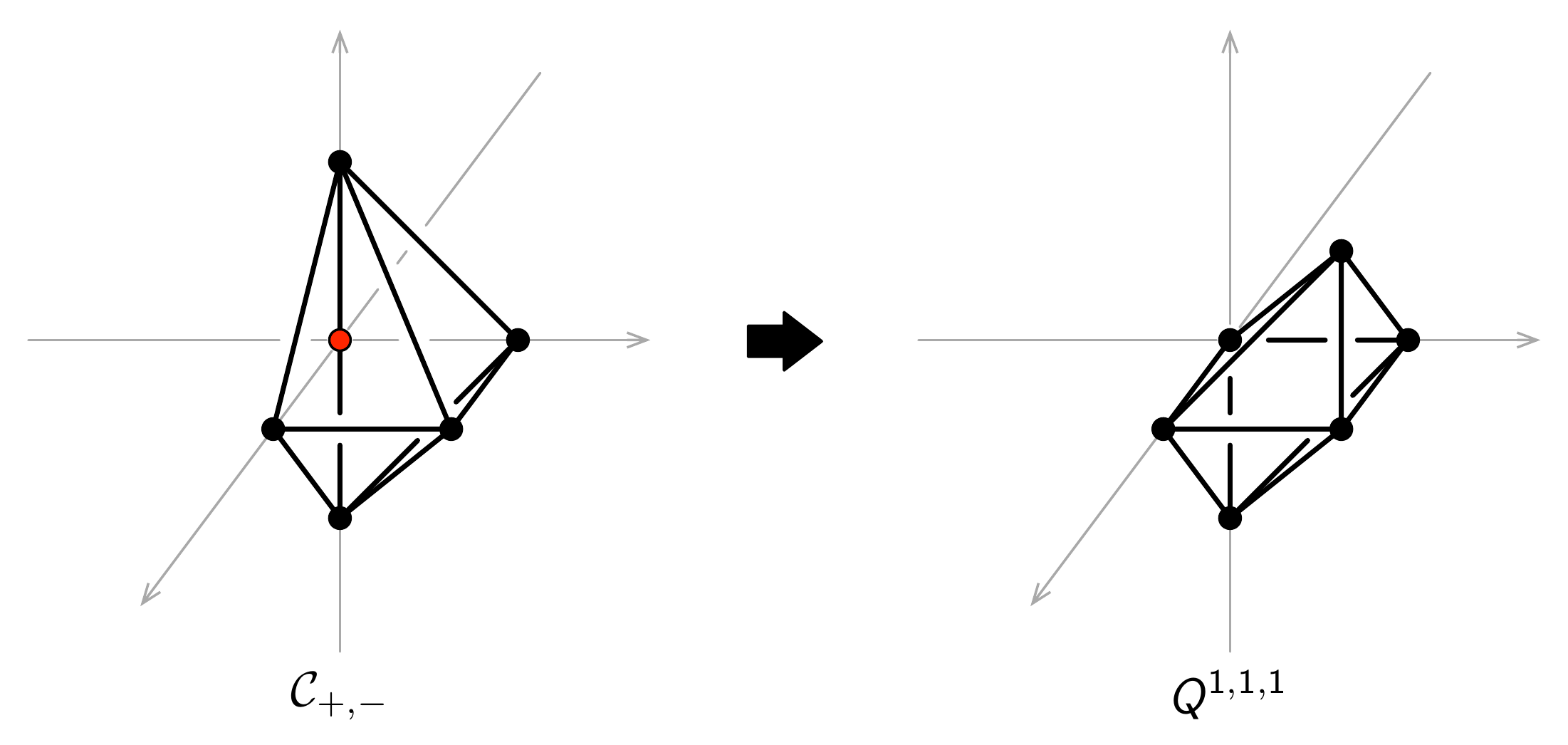}
\caption{Transition between the toric diagrams for $\mathcal{C}_{(+,-)}$ and $Q^{1,1,1}$.
\label{C+-_to_Q111}}
 \end{center}
 \end{figure}

We will now show that a $(0,2)$ KW deformation relates the gauge theories for $\mathcal{C}_{(+,-)}$ and $Q^{1,1,1}$. \fref{C+-_to_Q111} shows the corresponding toric diagrams.

\begin{figure}[H]
\begin{center}
\includegraphics[width=11cm]{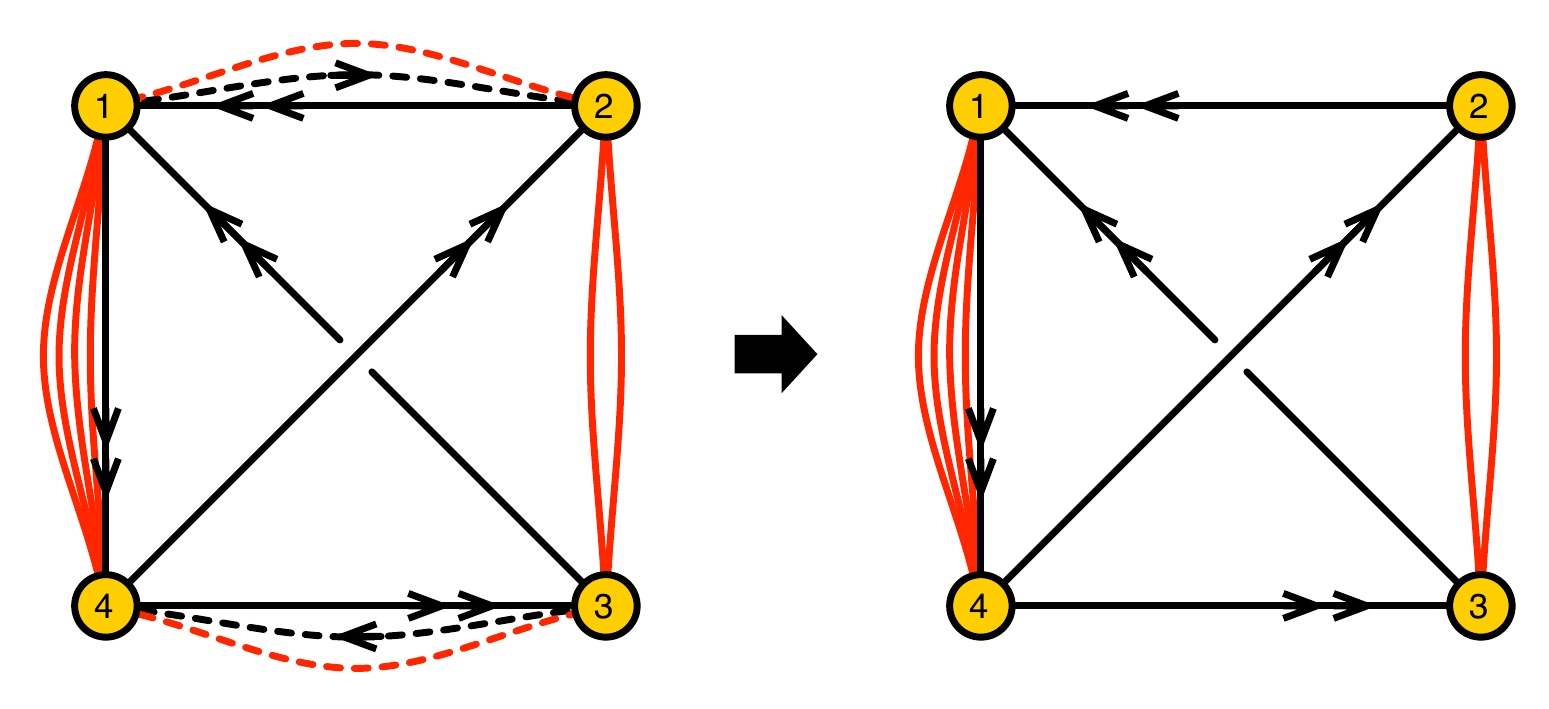}
\caption{Starting from the quiver for $\mathcal{C}_{(+,-)}$ and giving masses to the two chiral-Fermi pairs $(\Lambda_{12},Y_{12})$ and $(\Lambda_{34},Y_{34})$ (shown with dotted lines), we obtain the quiver for $Q^{1,1,1}$.
\label{quivers_KW}}
 \end{center}
 \end{figure}

In order to visualize the relation between the matter contents of both theories, it is convenient to consider their standard, instead of periodic quivers. These quivers are shown in \fref{quivers_KW}. The two gauge theories are connected by turning on appropriate mass terms for the $(\Lambda_{12},Y_{12})$ and $(\Lambda_{34},Y_{34})$ chiral-Fermi pairs and integrating them out. While the process is rather straightforward, we consider it is instructive to go over the computation in detail. Let us first consider the effect that integrating out these fields has on the quiver. As shown in \fref{quivers_KW}, the final quiver agrees with the one for $Q^{1,1,1}$, which was introduced in \cite{Franco:2015tna}.

This transformation of the quiver is promising. However, it is crucial to determine whether the $J$- and $E$-terms become those of $Q^{1,1,1}$. The mass terms correspond to adding a term in linear in $Y_{12}$ to the $E$-term for $\Lambda_{12}$ and a term linear in $Y_{34}$ to the $E$-term for $\Lambda_{34}$ as follows:\footnote{If we conjugate any of these two Fermis, the deformed term would be a $J$-term. The distinction between $J$- and $E$-terms is irrelevant in $(0,2)$ theories, which are symmetric under the exchange of any Fermi field with its conjugate. The terms to be deformed are univocally determined as those associated to the Fermi fields that acquire a mass and that have the same gauge quantum numbers as the massive chiral fields.}

\beq
\begin{array}{rcrcccrcc}
& & J \ \ \ \ \ \ \ \ \ \ \ \ \ \ \ \ \ & & & & E \ \ \ \ \ \ \ \ \ \ \ \ \ \ & & \\
 \Lambda_{12} : & \ \ \  & X_{21} \cdot Y_{12} \cdot Y_{21} - Y_{21} \cdot Y_{12} \cdot X_{21} & = & 0 & \ \ \ \  & \textcolor{blue}{-Y_{12}}+ V_{14} \cdot Q_{42} - X_{14} \cdot P_{42} & = & 0 \\ 
  \Lambda_{34} : & \ \ \  & X_{43} \cdot Y_{34} \cdot Y_{43} - Y_{43} \cdot Y_{34} \cdot X_{43} & = & 0 & \ \ \ \  &  \textcolor{blue}{Y_{34}}+P_{31} \cdot X_{14} - Q_{31} \cdot V_{14} & = & 0 \\ 
 \Lambda_{41}^{1} : & \ \ \  & Y_{12} \cdot Y_{21} \cdot X_{14} - X_{14} \cdot Y_{43} \cdot Y_{34} & = & 0 & \ \ \ \  & P_{42} \cdot X_{21} - X_{43} \cdot P_{31} & = & 0 \\ 
 \Lambda_{41}^{2} : & \ \ \  & X_{14} \cdot X_{43} \cdot Y_{34} - Y_{12} \cdot X_{21} \cdot X_{14} & = & 0 & \ \ \ \  & P_{42} \cdot Y_{21} - Y_{43} \cdot P_{31} & = & 0 \\ 
 \Lambda_{41}^{3} : & \ \ \  & Y_{12} \cdot Y_{21} \cdot V_{14} - V_{14} \cdot Y_{43} \cdot Y_{34} & = & 0 & \ \ \ \  & Q_{42} \cdot X_{21} - X_{43} \cdot Q_{31} & = & 0 \\ 
 \Lambda_{41}^{4} : & \ \ \  & V_{14} \cdot X_{43} \cdot Y_{34} - Y_{12} \cdot X_{21} \cdot V_{14} & = & 0 & \ \ \ \  & Q_{42} \cdot Y_{21} - Y_{43} \cdot Q_{31} & = & 0 \\ 
 \Lambda_{32}^{1} : & \ \ \  & Y_{21} \cdot X_{14} \cdot X_{43} - X_{21} \cdot X_{14} \cdot Y_{43} & = & 0 & \ \ \ \  & P_{31} \cdot Y_{12} - Y_{34} \cdot P_{42} & = & 0 \\ 
 \Lambda_{32}^{2} : & \ \ \  & Y_{21} \cdot V_{14} \cdot X_{43} - X_{21} \cdot V_{14} \cdot Y_{43} & = & 0 & \ \ \ \  & Q_{31} \cdot Y_{12} - Y_{34} \cdot Q_{42} & = & 0 
 \end{array} 
\label{deformed_E_J_C_+-}
 \eeq
 where we have indicated the new terms in blue. The relative minus sign is crucial for the deformation to have the desired effect. It is reminiscent of analogous relative signs between pairs of mass terms that are necessary in the original KW deformation \cite{Klebanov:1998hh} and its generalizations to other $4d$ theories \cite{Bianchi:2014qma}. $\Lambda_{12}$ and $\Lambda_{34}$ are integrated out and, consequently, the corresponding rows in \eref{deformed_E_J_C_+-} disappear. In addition, $Y_{12}$ and $Y_{34}$ are replaced by
\beq
 \begin{array}{ccl}
Y_{12} & = & V_{14} \cdot Q_{42} - X_{14} \cdot P_{42} \\[.1cm] 
Y_{34} & = & -(P_{31} \cdot X_{14} - Q_{31} \cdot V_{14})
 \end{array}
 \label{Y12_Y34}
 \eeq

Interestingly, whenever these two chiral fields appear in a surviving $J$- or $E$-term, they do so in pairs. This fact will be important for many cancellations, as we now show. Using \eref{deformed_E_J_C_+-} and \eref{Y12_Y34}, we obtain 
{\footnotesize
\beq
\begin{array}{rcl}
 \Lambda_{41}^{1} : & J:  & V_{14} \cdot Q_{42} \cdot Y_{21} \cdot X_{14} \textcolor{ForestGreen}{- X_{14} \cdot P_{42}\cdot Y_{21} \cdot X_{14} + X_{14} \cdot Y_{43} \cdot P_{31} \cdot X_{14}} -  X_{14} \cdot Y_{43} \cdot Q_{31} \cdot V_{14}  =  0 \\
& E: & P_{42} \cdot X_{21} - X_{43} \cdot P_{31}  =  0 \\[.1cm]
 \Lambda_{41}^{2} : & J: & \textcolor{ForestGreen}{-X_{14} \cdot X_{43} \cdot P_{31} \cdot X_{14}} + X_{14} \cdot X_{43} \cdot Q_{31} \cdot V_{14} - V_{14} \cdot Q_{42} \cdot X_{21} \cdot X_{14} \textcolor{ForestGreen}{+ X_{14} \cdot P_{42} \cdot X_{21} \cdot X_{14}} =  0 \\
 & E: & P_{42} \cdot Y_{21} - Y_{43} \cdot P_{31}  =  0 \\[.1cm] 
 \Lambda_{41}^{3} : & J:  & \textcolor{ForestGreen}{V_{14} \cdot Q_{42} \cdot Y_{21} \cdot V_{14}} - X_{14} \cdot P_{42} \cdot Y_{21} \cdot V_{14} + V_{14} \cdot Y_{43} \cdot P_{31} \cdot X_{14} \textcolor{ForestGreen}{- V_{14} \cdot Y_{43} \cdot Q_{31} \cdot V_{14}} = 0 \\
 & E: & Q_{42} \cdot X_{21} - X_{43} \cdot Q_{31}  =  0 \\[.1cm] 
 \Lambda_{41}^{4} : & J:  & - V_{14} \cdot X_{43} \cdot P_{31} \cdot X_{14} \textcolor{ForestGreen}{+ V_{14} \cdot X_{43} \cdot Q_{31} \cdot V_{14} - V_{14} \cdot Q_{42} \cdot X_{21} \cdot V_{14}} + X_{14} \cdot P_{42} \cdot X_{21} \cdot V_{14}  =  0 \\
 & E: & Q_{42} \cdot Y_{21} - Y_{43} \cdot Q_{31} = 0 \\[.1cm] 
 \Lambda_{32}^{1} : & J:  & -Y_{21} \cdot X_{14} \cdot X_{43} - X_{21} \cdot X_{14} \cdot Y_{43}  = 0 \\
  & E: &  \textcolor{BrickRed}{P_{31} \cdot V_{14} \cdot Q_{42} - P_{31} \cdot X_{14} \cdot P_{42}} + P_{31} \cdot X_{14} \cdot P_{42} - Q_{31} \cdot V_{14} \cdot P_{42} = 0 \\[.1cm] 
 \Lambda_{32}^{2} : & J:  & Y_{21} \cdot V_{14} \cdot X_{43} - X_{21} \cdot V_{14} \cdot Y_{43} = 0 \\
 & E: & \textcolor{BrickRed}{Q_{31} \cdot V_{14} \cdot Q_{42}} - Q_{31} \cdot X_{14} \cdot P_{42}  + P_{31} \cdot X_{14} \cdot Q_{42} \textcolor{BrickRed}{- Q_{31} \cdot V_{14} \cdot Q_{42}} = 0 
 \end{array}
 \eeq 
 }
For each $\Lambda_{34}^i$, the terms shown in green are identical to the $E$-terms of another $\Lambda_{34}^j$ and hence vanish on the moduli space. The pairs of terms shown in red vanish directly. Taking these cancellations into account, we get
 {\small 
 \beq
\begin{array}{rrcccrcc}
& J \ \ \ \ \ \ \ \ \ \ \ \ \ \ \ \ \ \ \ \ \ \ \  \ & & & & E \ \ \ \ \ \ \ \ \ \ \ \ \ \ \ & & \\
 \Lambda_{41}^{1} : & V_{14} \cdot Q_{42} \cdot Y_{21} \cdot X_{14} -  X_{14} \cdot Y_{43} \cdot Q_{31} \cdot V_{14}  & = & 0 & \ & P_{42} \cdot X_{21} - X_{43} \cdot P_{31}  & = & 0 \\
 \Lambda_{41}^{2} : &  X_{14} \cdot X_{43} \cdot Q_{31} \cdot V_{14} - V_{14} \cdot Q_{42} \cdot X_{21} \cdot X_{14} & = & 0 & & P_{42} \cdot Y_{21} - Y_{43} \cdot P_{31}  & = & 0 \\
 \Lambda_{41}^{3} : &  - X_{14} \cdot P_{42} \cdot Y_{21} \cdot V_{14} + V_{14} \cdot Y_{43} \cdot P_{31} \cdot X_{14} & = & 0 & & Q_{42} \cdot X_{21} - X_{43} \cdot Q_{31}  & = & 0 \\ 
 \Lambda_{41}^{4} : &  - V_{14} \cdot X_{43} \cdot P_{31} \cdot X_{14}+ X_{14} \cdot P_{42} \cdot X_{21} \cdot V_{14}  & = & 0 & & Q_{42} \cdot Y_{21} - Y_{43} \cdot Q_{31} & = & 0 \\
 \Lambda_{32}^{1} : &  -Y_{21} \cdot X_{14} \cdot X_{43} - X_{21} \cdot X_{14} \cdot Y_{43}  & = & 0 & \ &  P_{31} \cdot X_{14} \cdot P_{42} - Q_{31} \cdot V_{14} \cdot P_{42} & = & 0 \\
 \Lambda_{32}^{2} : &  Y_{21} \cdot V_{14} \cdot X_{43} - X_{21} \cdot V_{14} \cdot Y_{43} & = & 0 & & - Q_{31} \cdot X_{14} \cdot P_{42}  + P_{31} \cdot X_{14} \cdot Q_{42} & = & 0 
 \end{array} 
 \eeq 
 }
which are precisely the $J$- and $E$-terms for $Q^{1,1,1}$ \cite{Franco:2015tna}. We conclude that the mass deformation we introduced in \eref{deformed_E_J_C_+-} transforms $\mathcal{C}_{(+,-)}$ into $Q^{1,1,1}$, providing an explicit example of a $2d$ $(0,2)$ KW deformation.

\section{Conclusions}

\label{section_conclusions}

We introduced orbifold reduction, a novel method for generating $2d$ $(0,2)$ gauge theories associated to D1-branes probing toric CY 4-folds starting from $4d$ $\mathcal{N}=1$ gauge theories on D3-branes probing toric CY 3-folds. This procedure generalizes dimensional reduction and orbifolding. Orbifold reduction generates the periodic quiver on $T^3$ that encodes a $2d$ theory starting from the periodic quiver on $T^2$ for a $4d$ one. Equivalently, it generates brane brick models from brane tilings.

Orbifold reduction allows us to construct gauge theories for D1-branes probing toric CY 4-folds almost effortlessly. It is thus an ideal tool for generating new examples, which can in turn be used for studying both the D-brane configurations and the dynamics of the gauge theories. 

In order to illustrate the usefulness of orbifold reduction we presented three applications. We first showed how it connects $4d$ Seiberg duality to $2d$ triality. Next, we discussed how theories related by permutations of the sign vectors are automatically triality duals. The periodic quivers for such theories differ by a reorganization of layers that is similar to the motion of impurities in certain $4d$ toric theories due to Seiberg duality. Finally, we exploited orbifold reduction to construct an explicit example of a $2d$ $(0,2)$ KW deformation.

Our work suggests various interesting directions for future investigation. Let us mention a couple of them. It would be interesting to perform a systematic study of $2d$ KW deformations determining, among other things, the effect of the deformations on phase boundaries (the brane brick model analogues of zig-zag paths for brane tilings \cite{Franco:2015tya}) and establishing whether there is a general criterion that identifies theories with such deformations. Finally, it would be interesting to investigate whether a generalization of orbifold reduction similarly simplifies the construction of $0d$ $\mathcal{N}=1$ matrix models arising on the worldvolume of D(-1)-branes probing toric CY 5-folds. Understanding these theories is certainly a natural next step after brane tilings and brane brick models. Moreover, recent studies based on mirror symmetry suggest that these matrix models exhibit a quadrality symmetry \cite{Franco:2016qxh,quadrality}, making the development of new tools for studying such theories even more timely.

\acknowledgments

We would like to thank I. Klebanov and S.-T. Yau for useful and enjoyable discussions. We are also grateful to D. Ghim, A. Hasan and C. Vafa for collaboration on related topics. We gratefully acknowledge support from the Simons Center for Geometry and Physics, Stony Brook University, where some of the research for this paper was performed during the 2016 Simons Summer Workshop. R.-K. S. would also like to thank the hospitality of the CMSA at Harvard University where part of this project was completed. The work of S. F. is supported by the U.S. National Science Foundation grant PHY-1518967 and by a PSC-CUNY award. The work of S. L. was supported by the Samsung Science and Technology Foundation under Project Number SSTF-BA1402-08. The work of S. L. was also performed in part at the Institute for Advanced Study supported by the IBM Einstein Fellowship of the Institute for Advanced Study.

\newpage

\appendix 

\section{Additional Examples}

\label{section_appendix_examples}

\begin{figure}[ht]
\begin{center}
\includegraphics[height=5cm]{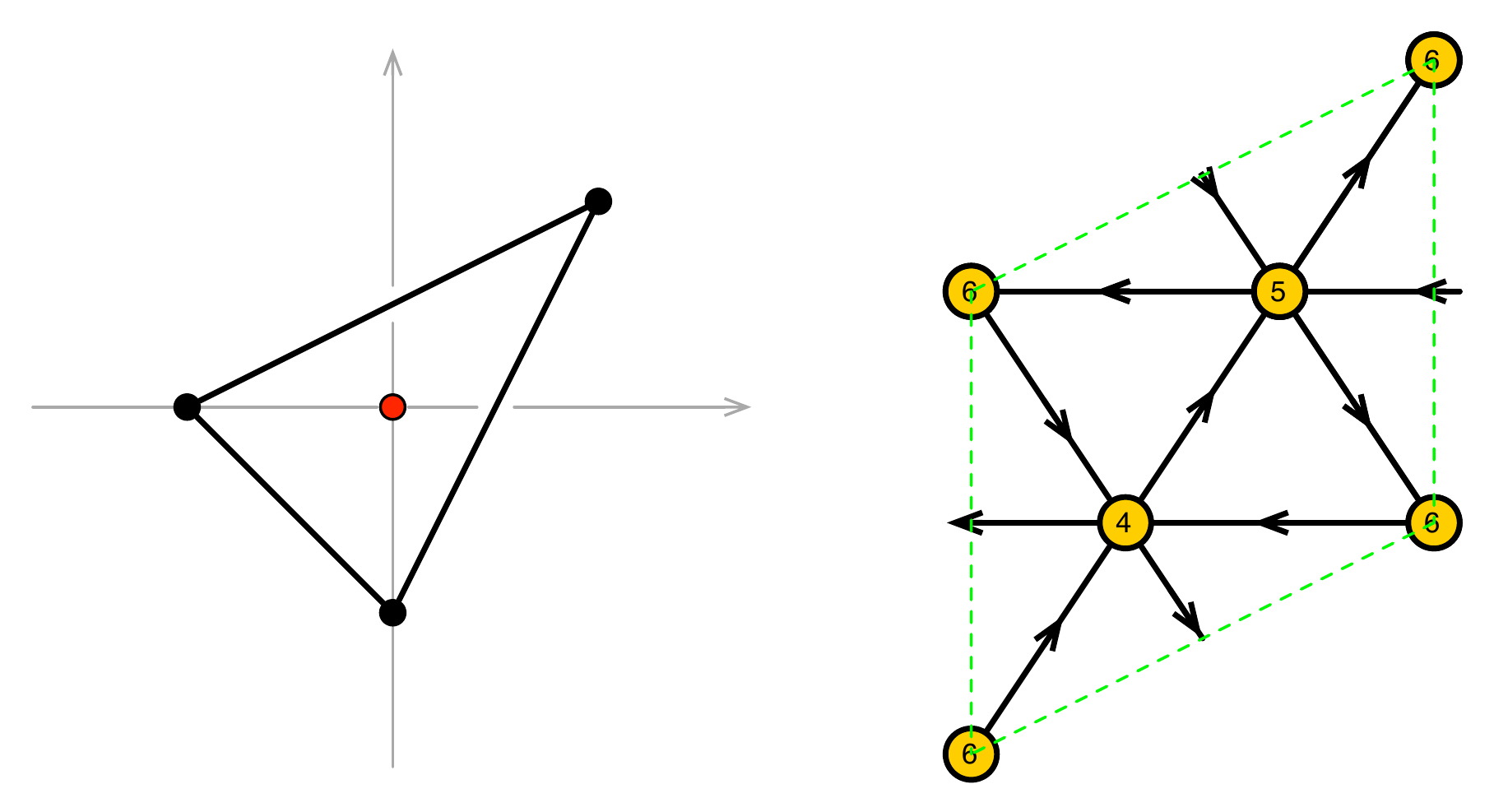}
\caption{Periodic quiver and toric diagram for $dP_0$.
\label{toric_quiver_dP0}}
 \end{center}
 \end{figure}

In this appendix we present two additional examples, obtained by orbifold reduction from the complex cone over $dP_0$, whose toric diagram and periodic quiver are shown in \fref{toric_quiver_dP0}.

\subsection{$M^{3,2}$}

\begin{figure}[H]
\begin{center}
\includegraphics[height=5cm]{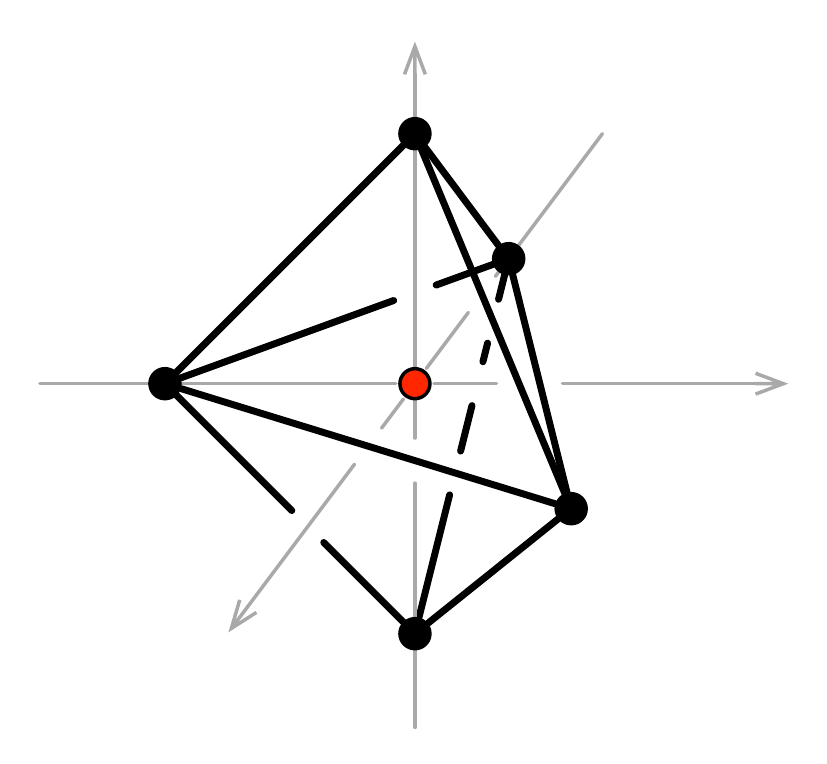}
\caption{Toric diagram for $M^{3,2}$. 
\label{toricm32}}
 \end{center}
 \end{figure}

Let us first consider an $s=(+,-)$ orbifold reduction of $dP_0$ using a $p_0$ associated to the central point of the toric diagram in \fref{toric_quiver_dP0}. The resulting theory corresponds to the toric diagram shown in \fref{toricm32}, which is usually referred to as $M^{3,2}$.

\begin{figure}[ht]
\begin{center}
\resizebox{.92\hsize}{!}{
\includegraphics[trim=0cm 0cm 0cm 0cm,totalheight=4.5 cm]{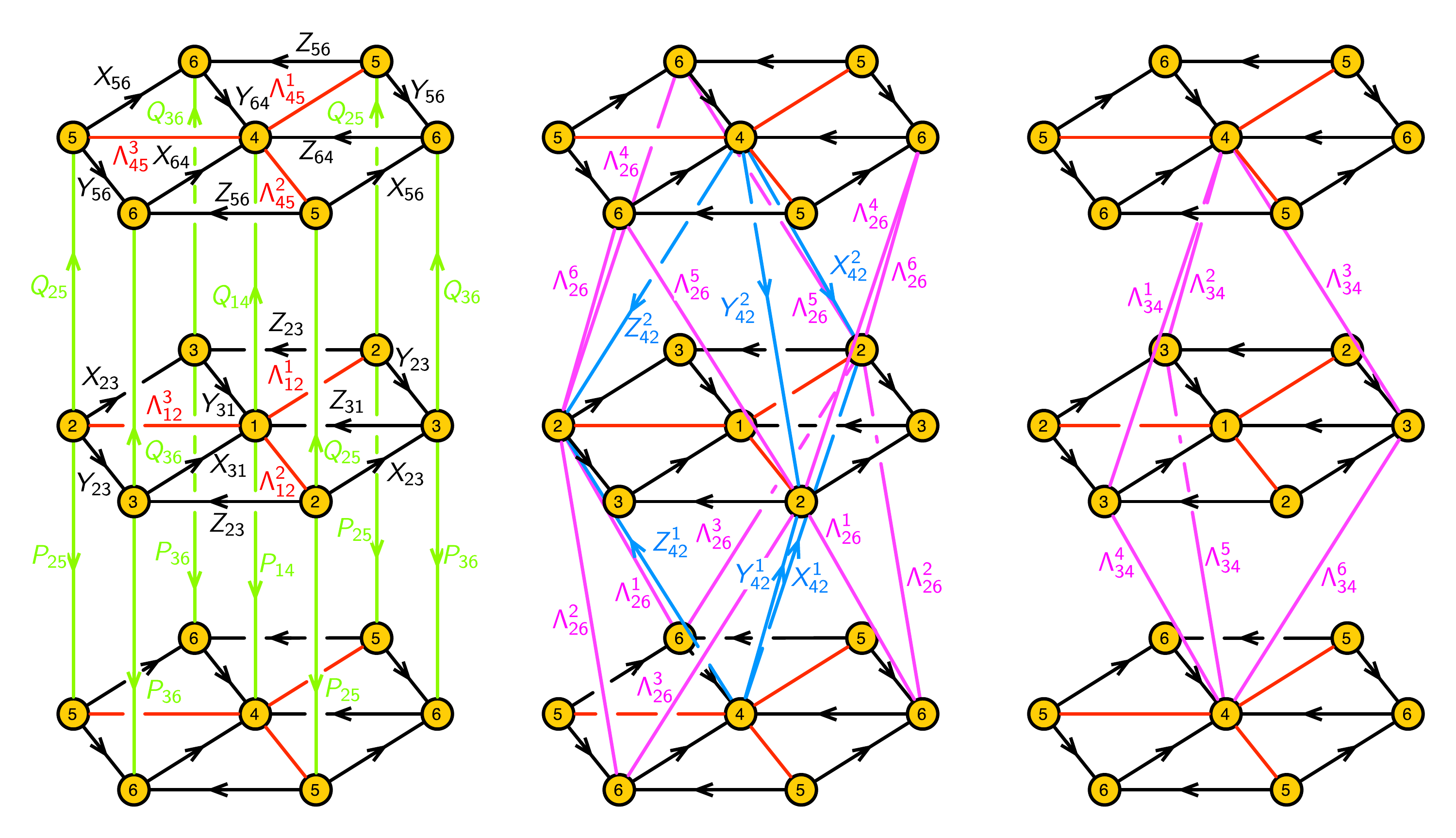}
}  
\vspace{-.2cm}\caption{Periodic quiver for $M^{3,2}$.
 \label{periodicquiverm32}}
 \end{center}
 \end{figure}

The periodic quiver for this theory is shown in \fref{periodicquiverm32}.

The $J$- and $E$- terms are:
{\small
\beq
\begin{array}{rclccclcc}
& & \ \ \ \ \ \ \ \ \ \ \ \ \ \ \ J & & & & \ \ \ \ \ \ \ \ \ \ \ \ \ \ E & & \\
\Lambda_{12}^{1} : & \ \ \  & Z_{23}\cdot Y_{31} - Y_{23}\cdot Z_{31} & = & 0 & \ \ \ \  & P_{14}\cdot X_{42}^{1} - Q_{14}\cdot X_{42}^{2} & = & 0 \\ 
\Lambda_{12}^{2} : & \ \ \  & X_{23}\cdot Z_{31} - Z_{23}\cdot X_{31} & = & 0 & \ \ \ \  & P_{14}\cdot Y_{42}^{1} - Q_{14}\cdot Y_{42}^{2} & = & 0 \\ 
\Lambda_{12}^{3} : & \ \ \  & Y_{23}\cdot X_{31} - X_{23}\cdot Y_{31} & = & 0 & \ \ \ \  & P_{14}\cdot Z_{42}^{1} - Q_{14}\cdot Z_{42}^{2} & = & 0 \\ 
\Lambda_{45}^{1} : & \ \ \  & Z_{56}\cdot Y_{64} - Y_{56}\cdot Z_{64} & = & 0 & \ \ \ \  & X_{42}^{1}\cdot P_{25} - X_{42}^{2}\cdot Q_{25} & = & 0 \\ 
\Lambda_{45}^{2} : & \ \ \  & X_{56}\cdot Z_{64} - Z_{56}\cdot X_{64} & = & 0 & \ \ \ \  & Y_{42}^{1}\cdot P_{25} - Y_{42}^{2}\cdot Q_{25} & = & 0 \\ 
\Lambda_{45}^{3} : & \ \ \  & Y_{56}\cdot X_{64} - X_{56}\cdot Y_{64} & = & 0 & \ \ \ \  & Z_{42}^{1}\cdot P_{25} - Z_{42}^{2}\cdot Q_{25} & = & 0 \\ 
\Lambda_{26}^{1} : & \ \ \  & Z_{64}\cdot Y_{42}^{1} - Y_{64}\cdot Z_{42}^{1} & = & 0 & \ \ \ \  & P_{25}\cdot X_{56} - X_{23}\cdot P_{36} & = & 0 \\ 
\Lambda_{26}^{2} : & \ \ \  & X_{64}\cdot Z_{42}^{1} - Z_{64}\cdot X_{42}^{1} & = & 0 & \ \ \ \  & P_{25}\cdot Y_{56} - Y_{23}\cdot P_{36} & = & 0 \\ 
\Lambda_{26}^{3} : & \ \ \  & Y_{64}\cdot X_{42}^{1} - X_{64}\cdot Y_{42}^{1} & = & 0 & \ \ \ \  & P_{25}\cdot Z_{56} - Z_{23}\cdot P_{36} & = & 0 \\ 
\Lambda_{26}^{4} : & \ \ \  & Z_{64}\cdot Y_{42}^{2} - Y_{64}\cdot Z_{42}^{2} & = & 0 & \ \ \ \  & Q_{25}\cdot X_{56} - X_{23}\cdot Q_{36} & = & 0 \\ 
\Lambda_{26}^{5} : & \ \ \  & X_{64}\cdot Z_{42}^{2} - Z_{64}\cdot X_{42}^{2} & = & 0 & \ \ \ \  & Q_{25}\cdot Y_{56} - Y_{23}\cdot Q_{36} & = & 0 \\ 
\Lambda_{26}^{6} : & \ \ \  & Y_{64}\cdot X_{42}^{2} - X_{64}\cdot Y_{42}^{2} & = & 0 & \ \ \ \  & Q_{25}\cdot Z_{56} - Z_{23}\cdot Q_{36} & = & 0 \\ 
\Lambda_{34}^{1} : & \ \ \  & Z_{42}^{2}\cdot Y_{23} - Y_{42}^{2}\cdot Z_{23} & = & 0 & \ \ \ \  & Q_{36}\cdot X_{64} - X_{31}\cdot Q_{14} & = & 0 \\ 
\Lambda_{34}^{2} : & \ \ \  & X_{42}^{2}\cdot Z_{23} - Z_{42}^{2}\cdot X_{23} & = & 0 & \ \ \ \  & Q_{36}\cdot Y_{64} - Y_{31}\cdot Q_{14} & = & 0 \\ 
\Lambda_{34}^{3} : & \ \ \  & Y_{42}^{2}\cdot X_{23} - X_{42}^{2}\cdot Y_{23} & = & 0 & \ \ \ \  & Q_{36}\cdot Z_{64} - Z_{31}\cdot Q_{14} & = & 0 \\ 
\Lambda_{34}^{4} : & \ \ \  & Z_{42}^{1}\cdot Y_{23} - Y_{42}^{1}\cdot Z_{23} & = & 0 & \ \ \ \  & P_{36}\cdot X_{64} - X_{31}\cdot P_{14} & = & 0 \\ 
\Lambda_{34}^{5} : & \ \ \  & X_{42}^{1}\cdot Z_{23} - Z_{42}^{1}\cdot X_{23} & = & 0 & \ \ \ \  & P_{36}\cdot Y_{64} - Y_{31}\cdot P_{14} & = & 0 \\ 
\Lambda_{34}^{6} : & \ \ \  & Y_{42}^{1}\cdot X_{23} - X_{42}^{1}\cdot Y_{23} & = & 0 & \ \ \ \  & P_{36}\cdot Z_{64} - Z_{31}\cdot P_{14} & = & 0 \\ 
 \end{array}
 \eeq
}

In \cite{Franco:2016qxh}, we constructed a different phase for $M^{3,2}$ using mirror symmetry. That theory is related to the one we have just presented by triality.

\subsection{$K^{3,2}$}

\begin{figure}[ht]
\begin{center}
\includegraphics[height=5cm]{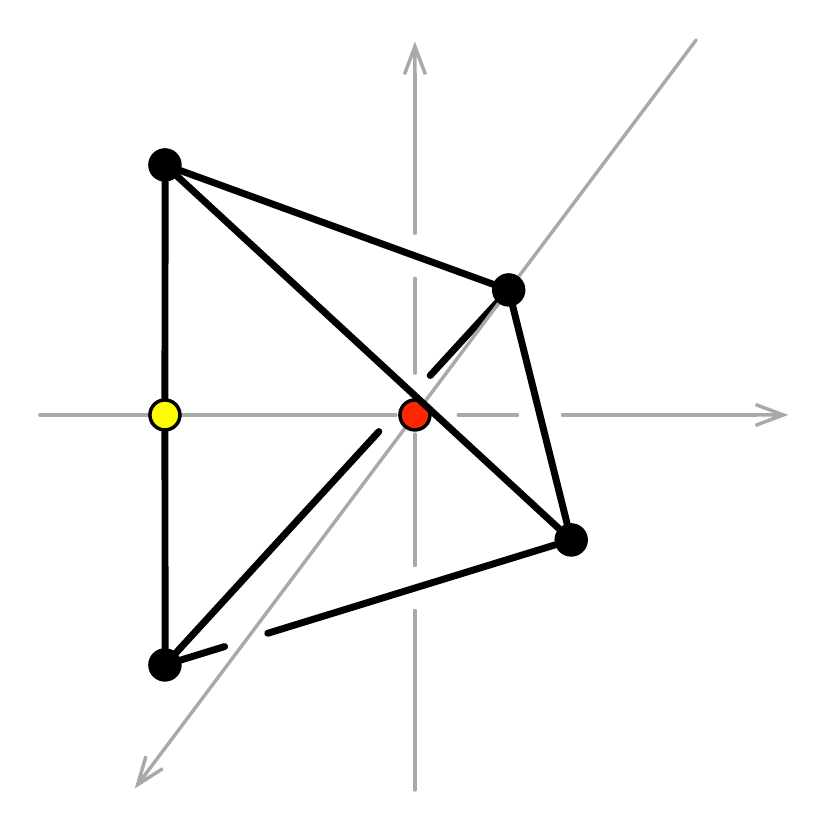}
\caption{Toric diagram for $K^{3,2}$. 
\label{toricK32}}
 \end{center}
 \end{figure}

Let us now construct another $s=(+,-)$ orbifold reduction of $dP_0$, but using a $p_0$ at one of the corners of the toric diagram in \fref{toric_quiver_dP0}. The resulting theory corresponds to the toric diagram shown in \fref{toricK32}. We will refer to this geometry as $K^{3,2}$.

\begin{figure}[H]
\begin{center}
\resizebox{.92\hsize}{!}{
\includegraphics[trim=0cm 0cm 0cm 0cm,totalheight=4.5 cm]{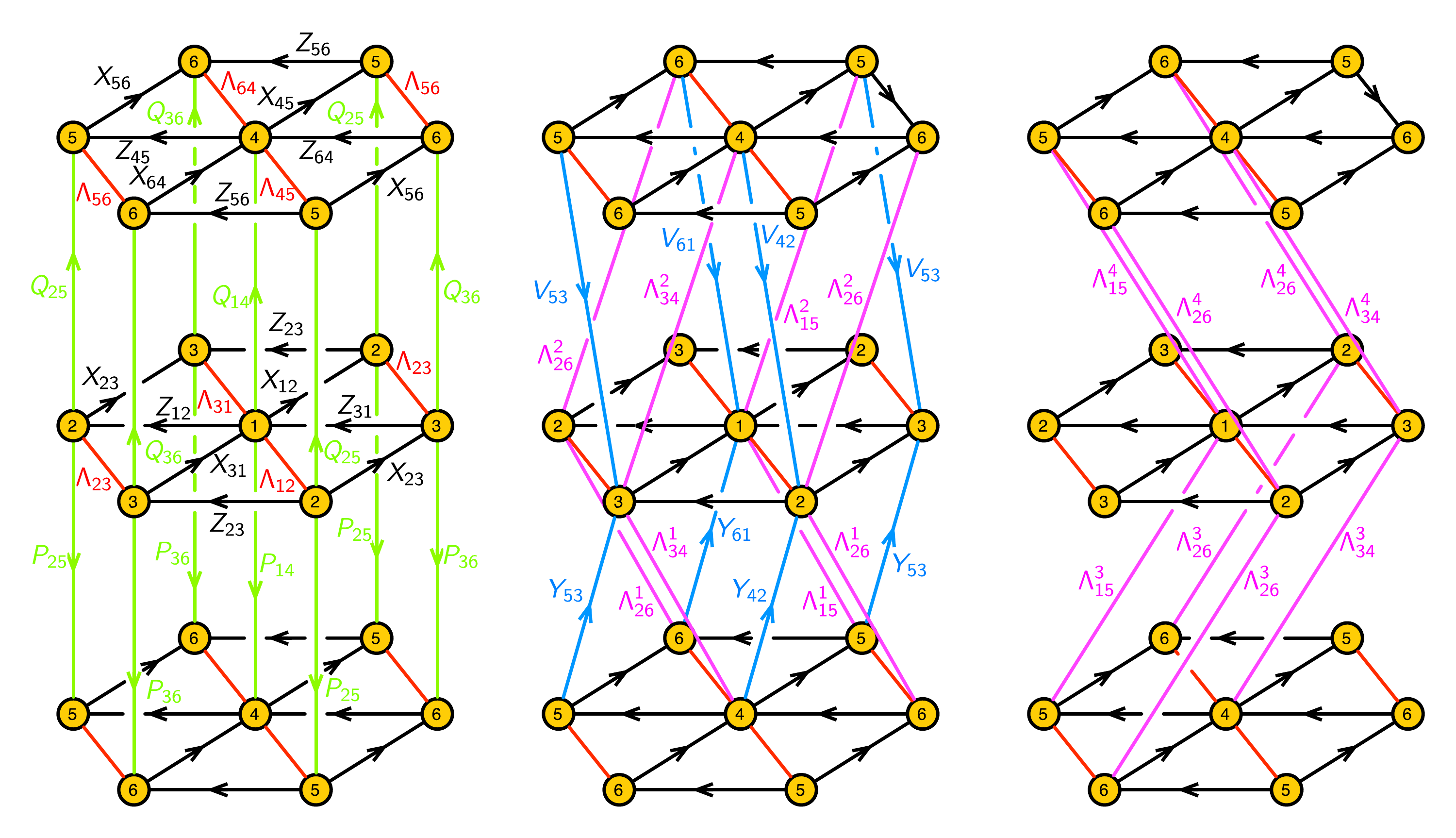}
}  
\vspace{-.2cm}\caption{Periodic quiver for $K^{3,2}$. 
\label{periodicquiverk32}
}
 \end{center}
 \end{figure}

\fref{periodicquiverk32} shows the periodic quiver for this theory.

Its $J$- and $E$-terms are:
{\small
\beq
\begin{array}{rclccclcc}
& & \ \ \ \ \ \ \ \ \ \ \ \ \ \ \ J & & & & \ \ \ \ \ \ \ \ \ \ \ \ \ \ E & & \\
\Lambda_{15}^{1} : & \ \ \  & Z_{56}\cdot Y_{61} - Y_{53}\cdot Z_{31} & = & 0 & \ \ \ \  & P_{14}\cdot X_{45} - X_{12}\cdot P_{25} & = & 0 \\ 
\Lambda_{26}^{1} : & \ \ \  & Z_{64}\cdot Y_{42} - Y_{61}\cdot Z_{12} & = & 0 & \ \ \ \  & P_{25}\cdot X_{56} - X_{23}\cdot P_{36} & = & 0 \\ 
\Lambda_{34}^{1} : & \ \ \  & Z_{45}\cdot Y_{53} - Y_{42}\cdot Z_{23} & = & 0 & \ \ \ \  & P_{36}\cdot X_{64} - X_{31}\cdot P_{14} & = & 0 \\ 
\Lambda_{15}^{2} : & \ \ \  & Z_{56}\cdot V_{61} - V_{53}\cdot Z_{31} & = & 0 & \ \ \ \  & Q_{14}\cdot X_{45} - X_{12}\cdot Q_{25} & = & 0 \\ 
\Lambda_{26}^{2} : & \ \ \  & Z_{64}\cdot V_{42} - V_{61}\cdot Z_{12} & = & 0 & \ \ \ \  & Q_{25}\cdot X_{56} - X_{23}\cdot Q_{36} & = & 0 \\ 
\Lambda_{34}^{2} : & \ \ \  & Z_{45}\cdot V_{53} - V_{42}\cdot Z_{23} & = & 0 & \ \ \ \  & Q_{36}\cdot X_{64} - X_{31}\cdot Q_{14} & = & 0 \\ 
\Lambda_{15}^{3} : & \ \ \  & Y_{53}\cdot X_{31} - X_{56}\cdot Y_{61} & = & 0 & \ \ \ \  & P_{14}\cdot Z_{45} - Z_{12}\cdot P_{25} & = & 0 \\ 
\Lambda_{26}^{3} : & \ \ \  & Y_{61}\cdot X_{12} - X_{64}\cdot Y_{42} & = & 0 & \ \ \ \  & P_{25}\cdot Z_{56} - Z_{23}\cdot P_{36} & = & 0 \\ 
\Lambda_{34}^{3} : & \ \ \  & Y_{42}\cdot X_{23} - X_{45}\cdot Y_{53} & = & 0 & \ \ \ \  & P_{36}\cdot Z_{64} - Z_{31}\cdot P_{14} & = & 0 \\ 
\Lambda_{15}^{4} : & \ \ \  & V_{53}\cdot X_{31} - X_{56}\cdot V_{61} & = & 0 & \ \ \ \  & Q_{14}\cdot Z_{45} - Z_{12}\cdot Q_{25} & = & 0 \\ 
\Lambda_{26}^{4} : & \ \ \  & V_{61}\cdot X_{12} - X_{64}\cdot V_{42} & = & 0 & \ \ \ \  & Q_{25}\cdot Z_{56} - Z_{23}\cdot Q_{36} & = & 0 \\ 
\Lambda_{34}^{4} : & \ \ \  & V_{42}\cdot X_{23} - X_{45}\cdot V_{53} & = & 0 & \ \ \ \  & Q_{36}\cdot Z_{64} - Z_{31}\cdot Q_{14} & = & 0 \\ 
\Lambda_{12} : & \ \ \  & X_{23}\cdot Z_{31} - Z_{23}\cdot X_{31} & = & 0 & \ \ \ \  & P_{14}\cdot Y_{42} - Q_{14}\cdot V_{42} & = & 0 \\ 
\Lambda_{23} : & \ \ \  & X_{31}\cdot Z_{12} - Z_{31}\cdot X_{12} & = & 0 & \ \ \ \  & P_{25}\cdot Y_{53} - Q_{25}\cdot V_{53} & = & 0 \\ 
\Lambda_{31} : & \ \ \  & X_{12}\cdot Z_{23} - Z_{12}\cdot X_{23} & = & 0 & \ \ \ \  & P_{36}\cdot Y_{61} - Q_{36}\cdot V_{61} & = & 0 \\ 
\Lambda_{45} : & \ \ \  & X_{56}\cdot Z_{64} - Z_{56}\cdot X_{64} & = & 0 & \ \ \ \  & Y_{42}\cdot P_{25} - V_{42}\cdot Q_{25} & = & 0 \\ 
\Lambda_{56} : & \ \ \  & X_{64}\cdot Z_{45} - Z_{64}\cdot X_{45} & = & 0 & \ \ \ \  & Y_{53}\cdot P_{36} - V_{53}\cdot Q_{36} & = & 0 \\ 
\Lambda_{64} : & \ \ \  & X_{45}\cdot Z_{56} - Z_{45}\cdot X_{56} & = & 0 & \ \ \ \  & Y_{61}\cdot P_{14} - V_{61}\cdot Q_{14} & = & 0 \\ 
 \end{array}
\eeq
}

\newpage

\bibliographystyle{JHEP}
\bibliography{mybib}


\end{document}